\documentclass[preprint, 11pt]{aastex}
\usepackage{lscape}
\usepackage{graphicx}
\usepackage{epstopdf}
\usepackage{epsfig}
\usepackage{natbib}

\newcommand{\degree}{$^{\circ}$}
\newcommand{\pmra}{$\mu_{\alpha}$}
\newcommand{\pmdec}{$\mu_{\delta}$}

\newcommand{\masyr}{mas\,yr$^{-1}$}
\newcommand{\kms}{km~s$^{-1}$}

\newcommand{\msun}{M$_{\sun}$}
\newcommand{\rsun}{R$_{\odot}$}
\newcommand{\mjup}{M$_{Jup}$}

\newcommand{\hi}{\ion{H}{1}}
\newcommand{\ci}{\ion{C}{1}}

\newcommand{\hei}{\ion{He}{1}}
\newcommand{\ki}{\ion{K}{1}}
\newcommand{\lii}{\ion{Li}{1}}
\newcommand{\mgi}{\ion{Mg}{1}}
\newcommand{\nii}{\ion{N}{2}}
\newcommand{\nai}{\ion{Na}{1}}
\newcommand{\caii}{\ion{Ca}{2}}
\newcommand{\oi}{\ion{O}{1}}
\newcommand{\oii}{\ion{O}{2}}
\newcommand{\oiii}{\ion{O}{3}}

\newcommand{\sii}{\ion{S}{2}}

\slugcomment{submitted to {\it AJ} on 2010 July 28; accepted on 2010 Aug 21}

\shorttitle{A Companion to TWA 30}
\shortauthors{Looper et al.}

\begin{document}

\title{A Widely-Separated, Highly-Occluded Companion to the Nearby Low-Mass T Tauri Star TWA 30\altaffilmark{1}}

\author{Dagny L. Looper\altaffilmark{2,3}, 
John J. Bochanski\altaffilmark{4}, 
Adam J. Burgasser\altaffilmark{3,4,5}, 
Subhanjoy Mohanty\altaffilmark{6},
Eric E. Mamajek\altaffilmark{7}, 
Jacqueline K. Faherty\altaffilmark{8,9},
Andrew A. West\altaffilmark{10},
Mark A. Pitts\altaffilmark{2,3}}

\altaffiltext{1}{This paper includes data gathered with the 6.5-m Magellan Telescopes located at Las Campanas Observatory, Chile.}

\altaffiltext{2}{Institute for Astronomy, University of
Hawai'i, 2680 Woodlawn Dr, Honolulu, HI 96822, USA; dagny@ifa.hawaii.edu} 

\altaffiltext{3}{Visiting Astronomer at the Infrared Telescope Facility, which
is operated by the University of Hawaii under Cooperative Agreement
no. NCC 5-538 with the National Aeronautics and Space Administration,
Office of Space Science, Planetary Astronomy Program} 

\altaffiltext{4}{MIT Kavli Institute for Astrophysics \& Space Research, 77 Massachusetts Ave, Building 37-664B, Cambridge, MA
02139, USA}

\altaffiltext{5}{Center for Astrophysics and Space Science, University of California San Diego, 9500 Gilman Drive, Mail Code 0424, La Jolla, CA 92093, USA}

\altaffiltext{6}{Imperial College London, 1010 Blackett Lab., Prince Consort Road, London SW7 2AZ, UK}

\altaffiltext{7}{Department of Physics and Astronomy, University of Rochester, P.O. Box 270171;
500 Wilson Boulevard, Rochester, NY 14627-0171, USA}

\altaffiltext{8}{Department of Physics and Astronomy, Stony Brook University, Stony Brook, NY 11794-3800, USA}

\altaffiltext{9}{Department of Astrophysics, American Museum of Natural History, Central Park West at 79th Street, New York, NY 10034, USA}

\altaffiltext{10}{Department of Astronomy, Boston University, 725 Commonwealth Ave, Boston, MA 02215, USA}

\begin{abstract}

We report the discovery of TWA 30B, a wide ($\sim$3400 AU), co-moving M dwarf companion to the nearby ($\sim$42~pc) young star TWA 30.  Companionship is confirmed from their statistically consistent proper motions and radial velocities, as well as a chance alignment probability of only 0.08\%.  Like TWA 30A, the spectrum of TWA 30B shows signatures of an actively accreting disk (\hi\ and alkali line emission) and forbidden emission lines tracing outflowing material ([\oi], [\oii], [\oiii], [\sii], and [\nii]).  We have also detected [\ci] emission in the optical data, marking the first such detection of this line in a pre-main sequence star.  Negligible radial velocity shifts in the emission lines relative to the stellar frame of rest ($\Delta$V $\lesssim$ 30 \kms) indicate that the outflows are viewed in the plane of the sky and that the corresponding circumstellar disk is viewed edge-on.  Indeed, TWA 30B appears to be heavily obscured by its disk, given that it is 5 magnitudes fainter than TWA 30A at K-band despite having a slightly earlier spectral type (M4 versus M5).  The near-infrared spectrum of TWA 30B also evinces an excess that varies on day timescales, with colors that follow classical T Tauri tracks as opposed to variable reddening (as is the case for TWA 30A).  Multi-epoch data show this excess to be well-modeled by a black body component with temperatures ranging from 630 to 880 K and emitting areas that scale inversely with the temperature.  The variable excess may arise from disk structure such as a rim or a warp at the inner disk edge located at a radial distance of $\sim$ 3--5 \rsun.   As the second and third closest actively accreting and outflowing stars to the Sun (after TWA 3), TWA 30AB presents an ideal system for detailed study of star and planetary formation processes at the low-mass end of the hydrogen-burning spectrum.
\end{abstract}


\section{Introduction}

The TW Hydrae Association (TWA) comprises the nearest population of classical T Tauri stars (cTTS), yet its membership remains incomplete.  Formally recognized as a young physical association composed of five stars by \cite{1997Sci...277...67K}, its membership swelled to 24 systems in little more than a decade (See \citealt{2005ApJ...634.1385M,2010ApJ...714...45L} and references there-in).  Many of these systems have been intensely studied because of their young age ($\sim$8 Myr; e.g., \citealt{1998ApJ...498..385S,1999ApJ...512L..63W,2006A&A...459..511B}) and close proximity to the Sun ($\sim$50 pc; \citealt{2005ApJ...634.1385M}).  The natal cloud from which these young stars sprung has since dispersed \citep{2009PASJ...61..585T}.   Mostly low-mass stars and brown dwarfs ($<$0.08 \msun) populate TWA, with only one object having a spectral type earlier than K5: the A0 V star TWA 11A.  TWA members have a variety of disk properties at a critical era in protoplanetary formation.  Therefore, identification of additional members provides invaluable case studies of disks and outflows of low-mass objects in our Galactic backyard.

Associations typically have an overdensity of stars against the background stellar population that assist in membership identification.  
However, the density of TWA members is roughly two orders of magnitude less than the local disk density: $\sim$0.002 \msun\ pc$^{-3}$ versus $\sim$0.1 \msun\ pc$^{-3}$ \citep{2007A&A...474..653V}.   The loose grouping of TWA members and their proximity to the Sun translate into a projected extent of $\sim$1000 square degrees on the sky.  Hence, the wide expanse of the TWA requires other methods for identifying new members.  Past techniques include identification by X-ray brightness in ROSAT All-Sky Survey data (RASS; \citealt{1999ApJ...512L..63W,2001ApJ...549L.233Z}), IR excess in the {\it IRAS} Point Source Catalog \citep{1989ApJ...343L..61D,1992AJ....103..549G}, and kinematic selections from the Tycho-2 catalog (\citealt{2001A&A...368..866M}, with spectroscopic confirmation by \citealt{2002A&A...385..862S}).  These surveys list only very bright objects, i.e., Tycho-2 is statistically complete down to only V=11.5 mag \citep{2000A&A...355L..27H}.  Therefore, such programs have been heavily biased against late-type members of the TWA.

Nevertheless, there have been several successful searches for substellar TWA members.   \cite{2002ApJ...575..484G} conducted a NIR photometric search over the expanse of TWA using the Two Micron All-Sky Survey Point Source Catalog (2MASS PSC, \citealt{2003yCat.2246....0C, 2006AJ....131.1163S}) and identified the two M8 substellar members 2MASS J1139$-$3159 (TWA 26) and 2MASS J1207$-$3932 (TWA 27).  High resolution imaging of TWA 27A  \citep{2004A&A...425L..29C,2005A&A...438L..25C} revealed a late-L type, low-mass companion ($\sim$8 \mjup; \citealt{2007ApJ...657.1064M}).  Previously, \cite{1999ApJ...512L..63W} performed a high-resolution imaging survey of known TWA members and found the M8 substellar member TWA 5B.  A proper motion survey of the Southern sky using the SuperCOSMOS Sky Survey \citep[SSS; ][]{2001MNRAS.326.1315H,2001MNRAS.326.1295H,2001MNRAS.326.1279H} uncovered the M8 substellar member SSSPM J1102$-$3159 (\citealt{2005A&A...430L..49S}; TWA 28).  Likewise, a near-infrared (NIR) photometric search for nearby, ultracool dwarfs using the DEep Near-Infrared Survey \citep[DENIS; ][]{1997Msngr..87...27E} by \cite{2007ApJ...669L..97L} uncovered the M9.5 substellar member DENIS J1245$-$4429 (TWA 29).

\cite{2010ApJ...714...45L} recently identified another new member of TWA from a survey which pairs red candidates in 2MASS PSC data with ROSAT X-ray catalog data -- the highly variable M5~$\pm$~1 pre-main sequence (PMS) star TWA 30.  In this paper we report the serendipitous discovery of a widely separated companion to TWA 30: 2MASS J11321822$-$3018316, which we term TWA 30B.   We describe our identification of this source and follow-up observations in Section 2, as well as additional spectroscopic coverage of TWA 30A.  In Section 3, we review our kinematic and spectroscopic analysis of TWA 30B.  In Section 4, we propose a toy model for the origins of the NIR excess emission in TWA 30B, compare the properties of TWA 30A and B, and discuss the possibility that this wide binary is a member of the nearby $\beta$ Pic Association.  We give our conclusions in Section 5.

\section{Observations}

\subsection{Discovery}

\begin{figure*}[htbp]
\centering
\includegraphics[scale=0.8]{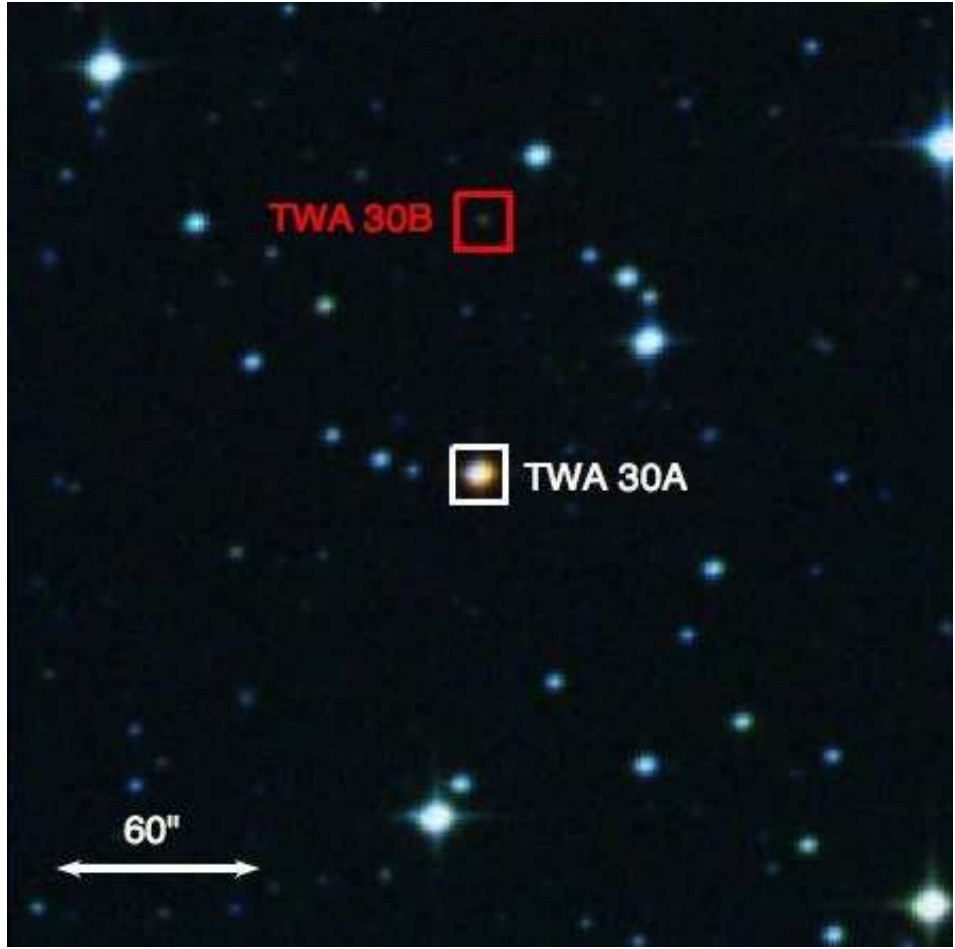}
\caption{Finder chart for TWA 30A (marked by a white box) and TWA 30B (marked by a red box).  The field is centered on TWA 30A and is 5 arcmin on a side, with north up and east to the left.  This image is a composite 3-color image made from DSS II $B$-, $R$-, and $I$-band images.}
\label{fig:finder}
\end{figure*}

We serendipitously discovered 2MASS J11321822$-$3018316 while making color finder charts for TWA 30A (see Figure \ref{fig:finder}).  2MASS PSC data showed this object as nearby ($\sim$80$\arcsec$) and red ($J-K_s\sim$1.6).  Using sky-survey plates from USNO-B1.0 and 2MASS (Table \ref{table:astrometry}), we found a similar proper motion for this candidate companion to TWA 30A (See Section 3.1).  Follow-up spectroscopy and astrometric measurements confirm its similar signatures of youth and shared kinematic properties.  From these multiple lines of evidence of companionship, we term this object TWA 30B.

\subsection{Optical Spectroscopy: Magellan/MagE}

\begin{figure*}[htbp]
\centering
\includegraphics[scale=0.8]{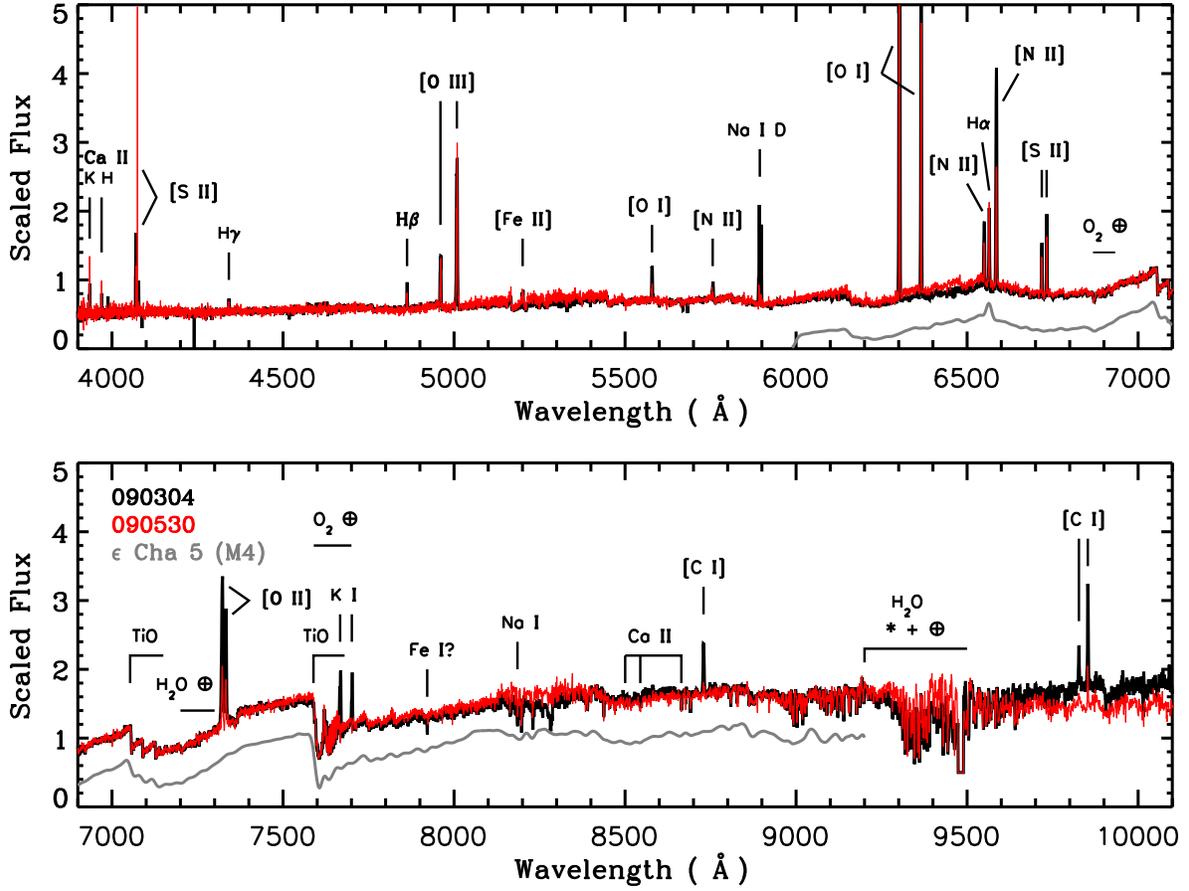}
\caption{The MagE spectra of TWA 30B from 2009 March 4 ({\it black}) and 2009 May 30 ({\it red}).  The young template RECX 5 ($\epsilon$ Cha 5, M4; \citealt{2004ApJ...609..917L}) is shown below these data in {\it gray} and provided the best spectral type match over 6000--8000 \AA.  All spectra have been normalized at 7500 \AA, and the template has been shifted down by 0.5 units for clarity.  The MagE spectra have a resolution of $R\sim$4100 while the template spectrum has a resolution of $R\sim$1000; however, the data were boxcar smoothed down to a comparable resolution for spectral classification.}
\label{fig:mage_spec}
\end{figure*}

TWA 30B was observed twice with the Magellan Echellete \citep[MagE;][]{2008SPIE.7014E.169M} on the 6.5 m Clay Telescope at Las Campanas Observatory, a cross-dispersed optical spectrograph which covers the $\sim$3000--10500 \AA\ passband in a single exposure.  Conditions on the first night were clear, while on the second night thin clouds were present (all observations are summarized in Table \ref{table:log}).  The 0$\farcs$7 slit was employed and aligned to parallactic for each observation, resulting in $R~\equiv~\lambda$ / $\Delta\lambda~\approx$~4100 with no binning.  A flux standard was observed during each run, and ThAr arc calibration images were obtained after each science exposure.  The spectrum was reduced with the MASE reduction tool pipeline \citep{2009PASP..121.1409B}.  The reduced spectra from both epochs are shown in Figure \ref{fig:mage_spec}.

\subsection{NIR Spectroscopy: IRTF/SpeX}

TWA 30B was observed over one year with 18 epochs of low-resolution near-infrared spectroscopy with the SpeX spectrograph \citep{2003PASP..115..362R} on the 3.0 m NASA Infrared Telescope Facility (IRTF).  The prism-dispersed mode of SpeX was used in conjunction with the 0$\farcs$5 slit, providing 0.7--2.5~$\micron$ spectroscopy at a mean resolution of $R\sim$150.  We also observed TWA 30B three times with the SpeX spectrograph in cross-dispersed mode (SXD) with the 0$\farcs$5 slit, resulting in a mean resolution of $R\sim$1200 over $\sim$0.8--2.4 $\micron$.  The details of these observations are recorded in Table \ref{table:log}.  Of note, all nights but 2010 January 24 UT and 2010 May 23 UT were clear and stable.  For telluric correction and flux calibration, observations of a nearby A0 V star were made immediately after each observation along with internal flat-field and argon arc lamp frames for calibrations.  All reductions were carried out with the Spextool package version 3.4 \citep{2004PASP..116..362C,2003PASP..115..389V} using standard settings.  

In addition, we observed TWA 30A over eight epochs spanning six months using the SpeX spectrograph in SXD mode with the 0$\farcs$5 slit (see Table \ref{table:log}).  This NIR data complements our previous spectroscopic coverage of TWA 30A (Looper et al. 2010) by extending the baseline coverage by 11 months.  Observations and reductions were carried out in the same manner as described above for TWA 30B.  

\subsection{Imaging}

TWA 30A was observed four times from 2008 December to 2010 January at the Cerro Tololo 4.0 m Blanco Telescope with the Infrared Side Port Imager (ISPI; \citealt{2004SPIE.5492.1582V}), using the $J$-band filter as part of an ongoing brown dwarf astrometric program (\citealt{2009AJ....137....1F,2010AJ....139..176F}).  While TWA 30A was the primary target, the large field of view (fov) of the detector ($\sim$10 arcmin fov with a 0$\farcs$3 pixel plate scale) allowed us to also derive positions for TWA 30B (equinox J2000; see Table \ref{table:astrometry}).  Images were taken in an identical manner: three images of 10 s exposures using three co-adds at each of the three dither positions.  At the beginning of the night, we obtained dark frames and domeflats.  Raw images were median-combined to produce sky frames, which were then subtracted from the raw data.  Otherwise, reduction procedures were based on the prescriptions put together by the ISPI team\footnote{See http://www.ctio.noao.edu/instruments/ir\_instruments/ispi/.}, utilizing a combination of routines from the Image Reduction and Analysis Facility (IRAF\footnote{IRAF is distributed by the National Optical Astronomy Observatory, which is operated by the Association of Universities for Research in Astronomy, Inc., under cooperative agreement with the National Science Foundation.}; \citealt{1986SPIE..627..733T}), as well as the publicly available software packages: WCSTOOLS and SWARP.  For further details, see \cite{2010ApJ...714...45L}.
\section{Analysis}

\subsection{Kinematics}

We determined a proper motion for TWA 30B of $\mu_{\alpha}=-$83~$\pm$~9 \masyr\ and $\mu_{\delta}=-$30~$\pm$~9 \masyr, using a weighted mean of the catalog positions of this object spanning over thirty years combined with our own measurements (Table \ref{table:astrometry}).   SSS lists an independent measurement of $\mu_{\alpha}=-$80~$\pm$~11 \masyr\ and $\mu_{\delta}=-$35~$\pm$~9 \masyr\ \citep{2001MNRAS.326.1315H}.  We adopted our calculated proper motion for analysis.  These  measurements are consistent within their respective errors to the proper motion measurement of TWA 30A, which has $\mu_{\alpha}=-$89.6~$\pm$~1.3 \masyr, $\mu_{\delta}=-$25.8~$\pm$~1.4 \masyr\ (UCAC3; \citealt{2010AJ....139.2200F}).  

Similarly, the motion of TWA 30B is consistent with TWA membership based on a convergent point analysis \citep{2005ApJ...634.1385M}.  Assuming a group vector of (\textit{U,V,W})=($-$10.2~$\pm$~0.5, $-$18.3~$\pm$~0.5, $-$4.9~$\pm$~0.5) for TWA (Mamajek, in prep), the proper motion of TWA 30B towards the convergent point is $\mu_{\upsilon}$~=~88.3~$\pm$~9.0 \masyr\ and the perpendicular proper motion is $\mu_{\tau}$~=~0.0~$\pm$~9.0 \masyr, i.e., perfectly consistent with motion toward the convergent point.  The predicted distance from this analysis is 44.0~$\pm$~4.9 pc from the Sun.  At its celestial position and estimated distance, TWA 30B resides only 5.4\degree\ or 9.7 pc from the centroid for the TWA ($\alpha$~=~174.2\degree, $\delta$~=~$-$35.6\degree, d~=~51.5~$\pm$~2.5 pc; Mamajek, in prep).  

We calculated the radial velocities (RV) of both epochs of the MagE spectra by cross-correlation with the M4 SDSS template from \cite{2007AJ....133..531B} using the \textit{xcorl} package in IDL \citep{2003ApJ...583..451M,2009ApJ...693.1283W}.  Our analysis covered the wavelength range: $\sim$6800--8900 \AA, a region that encompasses the TiO molecular bandhead that begins at 7056 \AA.  We find RVs of 10~$\pm$~4 \kms\ and 13~$\pm$~4 \kms\ for epochs 2009 March 4 and May 30, respectively, averaging to 12~$\pm$~3 \kms.  This is consistent with the predicted RV of a TWA member at the position of TWA 30B (11.0~$\pm$~1.0 \kms). 

Assuming the two components constitute a physical pair we derive properties of the system (see Table \ref{table:TWA30ABproperties}).  The weighted mean distance of the pair is 42~$\pm$~2 pc.  At an angular separation of 80$\farcs$2, this translates into a projected physical separation of $\rho$~=~3370~$\pm$~160 AU.  Statistically, the actual physical separation can be estimated as 1.26~$\times$~$\rho$~$\approx$~4250~$\pm$~200 AU \citep{1992ApJ...396..178F}.   Using the \cite{1998A&A...337..403B} models and assuming an age of 8 Myr, we estimate the physical parameters for TWA 30A (M5) as log T$_{eff}$~=~3.49, M~=~0.11 \msun\ and for TWA 30B (M4) as log T$_{eff}$~=~3.51, M~=~0.19 \msun.  Hence the binary's total mass is $\sim$0.30 \msun.  Assuming an actual semi-major axis of 4250 AU and a circular orbit, this yields r$_A$~=~1560 AU, v$_A$~=~0.09 \kms and r$_B$~=~2690 AU, v$_B$~=~0.16 \kms.  The period for this configuration is $\sim$0.5 Myr, translating into $\sim$16 orbits over the assumed age of the system.  

An interesting question arises -- is this system really bound or simply a chance alignment of two TWA members?  Assessing the boundedness for these systems presents difficulties as their predicted orbital velocities are on the order of $\sim$0.1 \kms, whereas our current astrometric accuracy is greater.  To calculate the chance alignment of two members, we generated a Monte Carlo simulation of 10,000 fake TWA clusters given the following distributions of RA, Dec, and distance with normal deviates generated for each parameter: (RA~=~174.6\degree, 1$\sigma$ dispersion~=~11.9\degree), (Dec~=~$-$36.3\degree, 1$\sigma$ dispersion~=~6.3\degree), (distance~=~54.7 pc, 1$\sigma$ dispersion~=~12.5 pc).  The mean distance and dispersion are based on revised cluster parallax distances to the individual members and are calculated based on the best available proper motions and a new space velocity and convergent point solution (RA~=~99.8\degree, Dec~=~$-$27.7\degree, S$_{tot}$~=~22.0 \kms).  We find the median minimum separation between two TWA members as 2816$\arcsec$, with a 95\% confidence level of 520--6433\arcsec and a 68\% confidence level of 1423--4548\arcsec.  In only 8 simulations out of 10,000 was the minimum separation between two TWA members $<$80\arcsec.  We therefore estimate a 0.08\% probability of a chance alignment for the components of TWA 30AB.        

Of note, TWA contains at least one other proposed very wide binary -- TWA 1 and 28, which has an angular separation of 735.3$\arcsec$ or a projected physical separation of 40,440 AU at a distance of 55 pc.  The co-motion between TWA 1 and 28 is 1.0~$\pm$~1.9 \masyr, which translates into an orbital motion of 0.3~$\pm$~0.7 \kms.  In only 4.74\% of our simulations is the minimum separation between fake TWA members less than 735.3$\arcsec$.  The astrometric data for TWA 1/28 provides further support for the pair to be co-distant and co-moving.  The \cite{2007A&A...474..653V} astrometry for TWA 1 and the \cite{2009A&A...503..281T} astrometry for TWA 28 agree within 1.0~$\pm$~1.9 \masyr\ in $\mu_{\alpha}$ and 0.1~$\pm$~1.6 \masyr\ in $\mu_{\delta}$.  This translates into a difference of 0.27~$\pm$~0.66 \kms\ if the pair are co-distant.  If they share velocity but not distance, this places an upper limit on the difference in their distance at 3.1 pc (3$\sigma$).  These simulations suggest (along with precise astrometry for TWA 1 and 28) that both the TWA 30AB and TWA 1/28 systems are physical binaries.

\subsection{Optical Spectroscopy}

\subsubsection{Spectral Morphology}

We classified TWA 30B by comparison to M3--M5.75 $\eta$ Cha templates (denoted by RECX \#) from \cite{2004ApJ...609..917L}.  $\eta$ Cha has a similar age ($\sim$7 Myr; \citealt{1999ApJ...516L..77M,2001MNRAS.321...57L}) to TWA; therefore, the templates should provide a better continuum fit with comparable surface gravities than would comparison to field templates, which have higher surface gravities.  The TWA 30B spectra were boxcar smoothed down to comparable resolution of the templates ($R\sim$1000) and then both data and template were normalized at 7500 \AA.  The best match to the template spectrum over the wavelength range 6000--8000 \AA\ was determined by eye.  This wavelength range covers many broadband features typically used to classify M dwarfs such as TiO, VO, and CrH \citep{1991ApJS...77..417K}.   

The best fit to both epochs was to the M4 template RECX 5.  We show the two unsmoothed epochs of MagE data along with the RECX 5 template in Figure \ref{fig:mage_spec}.  The continuum of TWA 30B shows little to no variation between epochs and matches well to the M4 template without having to account for variable reddening as is the case with TWA 30A (Looper et al. 2010).  We therefore adopt an optical spectral type of M4~$\pm$~0.5 for TWA 30B.

\subsubsection{Emission and Absorption Lines}

\begin{figure*}[htbp]
\centering
\includegraphics[scale=0.85]{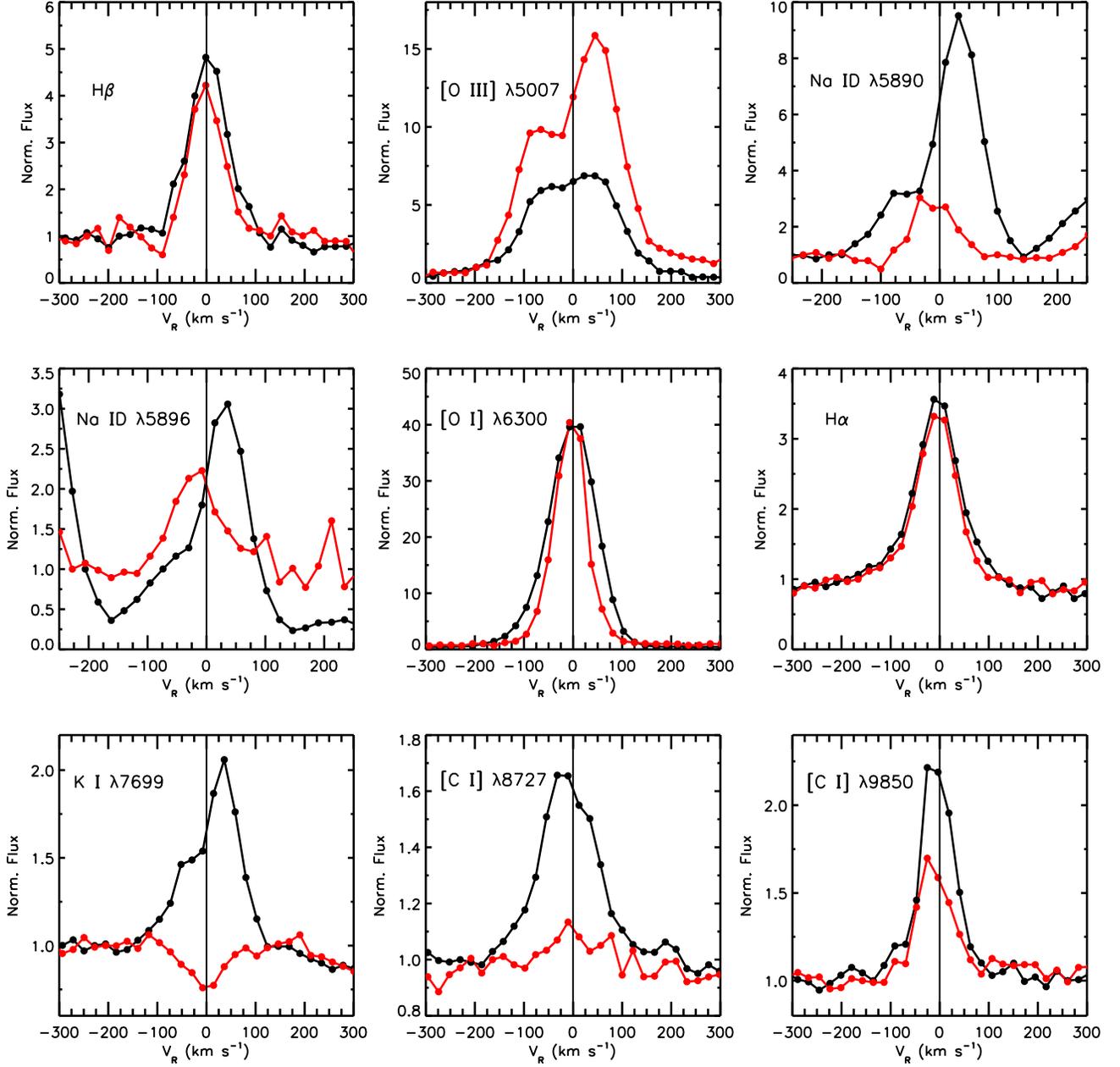}
\caption{The radial velocity profiles of several prominent features in the two MagE spectra of TWA 30B, color-coded identically to that of Figure \ref{fig:mage_spec}.  For each panel, the flux has been normalized at $-$250 \kms\ and the star's rest velocity (+12 \kms) has been subtracted out.  Each feature is labeled in the inset.}
\label{fig:rv_plot}
\end{figure*}

The continuum of TWA 30B varies little between the two epochs of MagE data.  However, the forbidden emission lines (FELs): [\sii], [\oi], [\oii], [\oiii], [\nii], and [\ci]; \nai D emission; and \lii\ and \ki\ absorption are variable in both line strength and profile shape (Table \ref{table:EWs} and Figures \ref{fig:mage_spec} \& \ref{fig:rv_plot}), while the Balmer emission lines are roughly constant ($\Delta$EW~$<$~15\%).

Like its companion, TWA 30B has FELs of [\sii], [\oi], and [\nii] -- signs of outflow and accretion in cTTS \citep{2003ApJ...592..266M}.  Notably, these lines are an order of magnitude stronger in the spectrum of TWA 30B than in TWA 30A.  Particularly [\nii] $\lambda$6547, 6583, which are just on the threshold of detection in TWA 30A (EWs~=~$-$0.1 to $-$0.4 and $-$0.3 to $-$1.0 \AA, respectively) are enhanced to well-detectable levels in TWA 30B (EWs~=~$-$3.0 to $-$7.2 and $-$10.8 to $-$25.1 \AA, respectively).  The strength of [\oi] $\lambda$6300 is increased from EWs of $-$3.1 to $-$15 \AA\ in TWA 30A to EWs of $-$85 to $-$226 \AA\ in TWA 30B, while the strength of [\oiii] $\lambda$5007 is twice as strong in the spectrum of the latter source.  This is a high excitation line that requires extremely high temperatures and low densities.   

All of the emission lines, with the exception of \caii\ H+K, are stronger in the 2009 March 4 spectrum of TWA 30B.  The FELs increase by a factor of $\sim$2--3 with the exception of [\oiii] $\lambda$4958, 5007 emission lines, which are fairly stable ($\Delta$EWs~=~0--30\%; Table \ref{table:EWs}).  The gravity sensitive \ki\ lines at 7665, 7699 \AA\ go into $\it{emission}$ in the spectrum of 2009 March 4 while they are in absorption in the spectrum of 2009 May 30.  This is also analogous to the changes seen in this line in TWA 30A.  The \nai D $\lambda$5890  feature in the 2009 May 30 data has a P Cygni profile, indicating outflowing winds along the line of sight.  The H$\alpha$ and H$\beta$ lines are nearly constant in both line EWs ($-$7 \AA) and symmetric profiles in the two epochs.  The full-width at 10\% peak intensity of H$\alpha$ is 220 \kms\ and 200 \kms\ for the 2009 March 4 and 2009 May 30 data, respectively.  The width of this line meets the $\ge$200 \kms\ threshold for accretion proscribed by \cite{2003ApJ...582.1109W}.  However, it is important to note that we expect this line to be both weakened and symmetrical if the disk is aligned edge-on, as is the case for MHO-5 in Taurus \citep{2005ApJ...626..498M}.

Another gravity sensitive line, \lii\ $\lambda$6707, which is also used as a mass/age indicator \citep{1997ApJ...482..442B}, has a comparable strength (EW~=~0.46~$\pm$~0.13 \AA) to other TWA M members (0.40 to 0.65 \AA, \citealt{2004ARA&A..42..685Z}) in the 2009 May 30 data but is not detected in the 2009 March 4 data.  Together with the stronger emission line strengths in the spectrum of 2009 March 4, this indicates that the photospheric lines of \lii\ and \ki\ are filled in (veiled) by enhanced accretion during this date.  In Figure \ref{fig:rv_plot}, we have plotted the radial velocity profiles of many of these lines after subtracting the radial velocity of TWA 30A (+12 \kms).  The centers of the photospheric lines \ki\ $\lambda$7699, H$\alpha$ and H$\beta$ match this stellar rest velocity ($\Delta~v~\lesssim$~5 \kms).  All other lines (FELs, emission, and absorption) show only small deviations from the stellar rest frame ($\Delta~v$~$<$~30 \kms; Table \ref{table:EWs}), suggesting that associated outflows are observed in the plane of the sky, so that the (assumed perpendicular) circumstellar disk is viewed edge-on.  

TWA 30B shows no \hei\ emission at 4471, 5876, 6678, and 7065 \AA.  \hei\ is present in the accreting late-M members TWA 27AB and TWA 28 and weakly in emission in the accreting late-K TWA 1 and the early-M TWA 3 \citep{2009ApJ...696.1589H}.  On the other hand, the spectrum of TWA 30B exhibits [\ci] $\lambda$8727, 9824, and 9850 in emission in both epochs, which is not present in the spectrum of TWA 30A or any known PMS stars.  \cite{2009A&A...496..725E} have suggested the use of this line in combination with the fine structure MIR lines of [\ci] to constrain the size of gaseous protoplanetary disks.  Their model for a T$_{eff}$~=~4000 K, log g~=~3.5, R~=~2.5 \rsun\ star predicted EWs of 2--5 m\AA\ for the NIR [\ci] lines.  We have measured EWs of $-$0.4 to $-$3.4 \AA: a factor of 1000 times greater in strength.  

\subsection{NIR Spectroscopy}

\subsubsection{Spectral Morphology}

\begin{figure*}[htbp]
\centering
\includegraphics[scale=0.55]{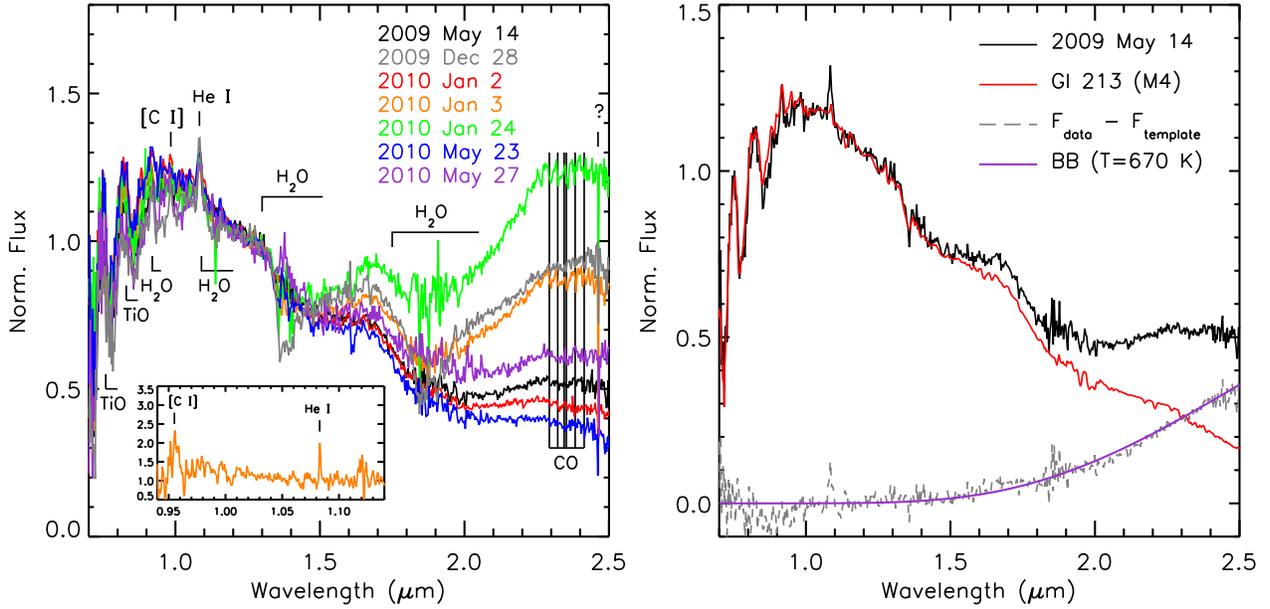}
\caption{{\bf Left}: The SpeX prism spectra of TWA 30B at the stated epochs.  All spectra have been normalized at 1.27 $\mu$m.  The spectrophotometry and colors calculated from these data are listed in Table \ref{table:specphot}.  The inset shows the SXD data of TWA 30B from 2010 Jan 3 ({\it yellow}) over the region encompassing the NIR \hei\ and [\ci] emission features.  {\bf Right}: The spectrum of TWA 30B from 2009 May 14 ({\it black}) in comparison to the field standard Gl 51 ({\it red}); both spectra have been normalized at 1.27 $\mu$m.  The subtracted flux of the template (Gl 51) from the data (TWA 30B) is shown as a  {\it dashed gray} line.  We have fit a blackbody with T$_{eff}$=670 K ({\it purple}) to the subtracted flux.}
\label{fig:all_nir}
\end{figure*}

In the left panel of Figure \ref{fig:all_nir}, we show seven NIR spectra of the 18 epochs, which span the full range of colors and dates observed.  Since we did not have a set of young M dwarfs observed at similar wavelengths and resolution as our data, we classified the TWA 30B NIR spectra by comparison to field M dwarfs taken with an identical set-up (Kirkpatrick et al.\ 2010): Gl 752A (M3V), Gl 213 (M4 V), Gl 51 (M5V), LHS 1375 (M6 V), and vB8 (M7 V).  In the right panel of Figure \ref{fig:all_nir}, we show the first epoch of our TWA 30B NIR data from 2009 May 14 UT in comparison to the M4 standard Gl 213.  The strengths of the TiO and H$_2$O bands as well as the overall morphology from 0.7--1.3 $\micron$ match well between the standard and data for all 18 epochs.  This region is similar to the wavelength coverage (0.9--1.4 $\micron$) proscribed by Kirkpatrick et al.\ (2010) for NIR spectral classification.  At longer wavelengths ($H$ and $K$), the spectra of TWA 30B showed variable levels of NIR excess.  We therefore adopt a NIR spectral type of M4~$\pm$~0.5.

\begin{figure*}[htbp]
\centering
\includegraphics[scale=0.8]{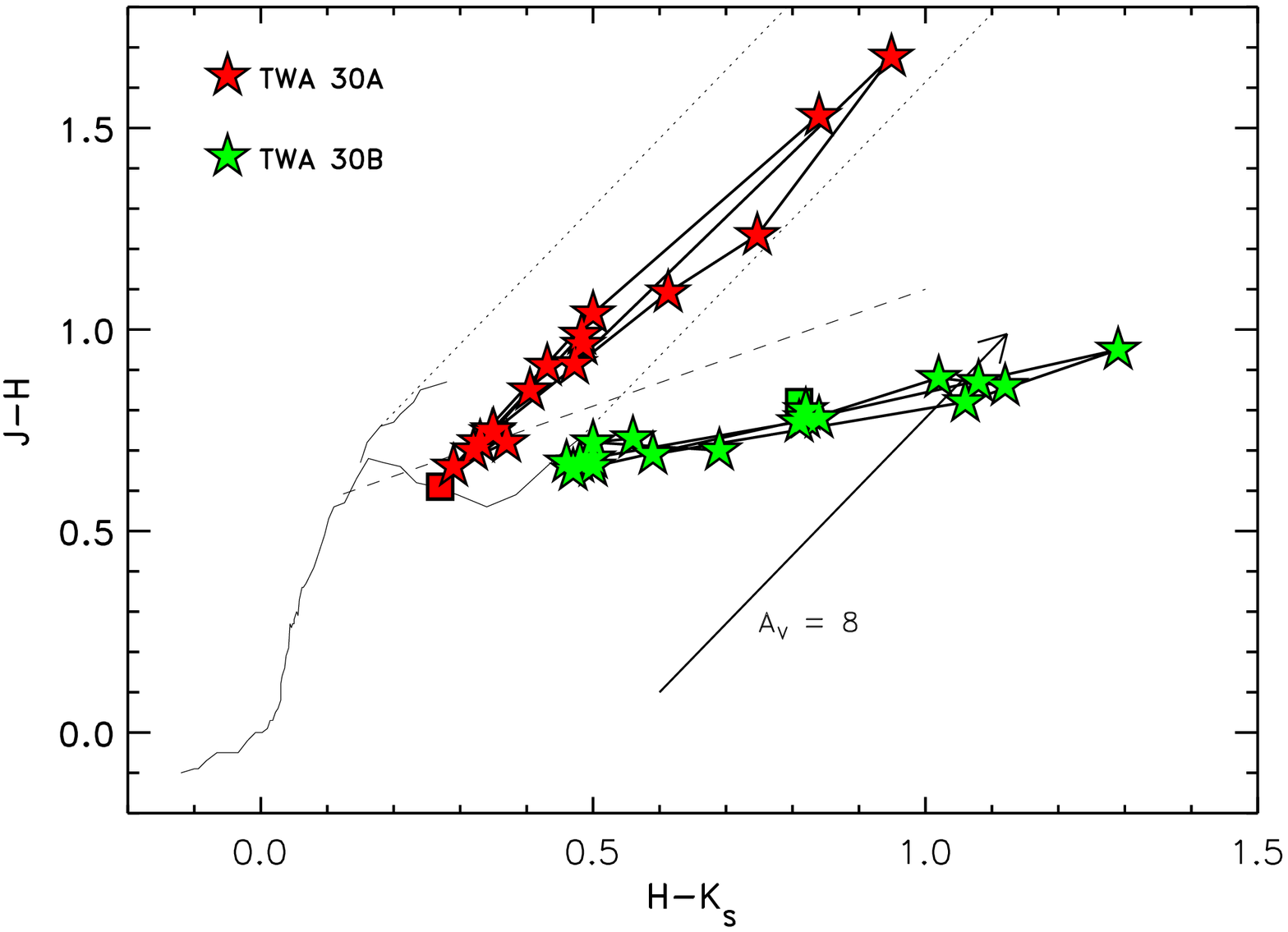}
\caption{Near-infrared colors of TWA 30A ({\it red five pointed stars}) and TWA 30B ({\it green five pointed stars}) derived from our multi-epoch SpeX data, plotted along with the single epoch of 2MASS data available for each source ({\it squares}).  The dwarf track ($\ge$M0 from \citealt{1992ApJS...82..351L}; earlier types from \citealt{1995ApJS..101..117K}) is the lower solid line and the giant track \citep{1988PASP..100.1134B} is the upper solid line, the color-color space occupied by reddened dwarfs is outlined by two parallel dotted tracks, the cTTS locus \citep{1997AJ....114..288M} is shown as a single dashed line, and the vector at the lower right indicates a reddening vector of A$_V$=8 mag.  Both the dwarf and giant tracks have been transformed to the CIT photometric system, which is similar to the 2MASS system \citep{2001AJ....121.2851C}.  Each set of stars is connected by a solid straight line to delineate the sequence of observations.  Note that the TWA 30A track is parallel to the reddening vector while the TWA 30B track is roughly parallel to the cTTS locus.}
\label{fig:jhk}
\end{figure*}

We measured the spectrophotometry of these data using the 2MASS $JHK_s$ filter profiles and report these values in Table \ref{table:specphot} and plot them in $JHK_s$ color-color space (along with TWA 30A) in Figure \ref{fig:jhk}.   These  epochs have $J-K_s$ measurements of 1.1--2.2 mag, much redder than the 0.86~$\pm$~0.16 mag average of 2126 field M4 dwarfs in the SDSS DR5 sample with 2MASS colors \citep{2008AJ....135..785W}.  The original 2MASS $J-K_s$ color is 1.63~$\pm$~0.06 mag.  TWA 30B therefore shows a variable amount of NIR excess, with a sharp increase from 2009 December 28 UT ($J-K_s$~=~1.3 mag) to the 2010 January 2--3 UT epochs ($J-K_s$~=~1.9 to 2.2 mag) before settling down to a more modest NIR excess in the 2010 January 24--28 UT data ($J-K_s$~=~1.1 to 1.2 mag).   Finally in the 2010 May data, the NIR excess increases again ($J-K_s$~=~1.6 to 2.0 mag).  Graphically, we see that this NIR color variation in TWA 30B is nearly parallel to the cTTS locus \citep{1997AJ....114..288M} in Figure \ref{fig:jhk}, while the NIR variability in TWA 30A is parallel to the reddening vector from \cite{1989ApJ...345..245C}.  The slope differences in color tracks suggests that the excesses in the these two sources arise from different origins or viewing geometries.  

\subsubsection{Emission and Absorption Lines}

Most of the spectroscopic NIR data we obtained for TWA 30B is too low resolution ($R\sim$150) to examine the emission and absorption line features.  We note, however, that even at these resolutions strong \hei\ emission appears at 1.083 $\mu$m and blended [\ci] emission at 0.982 and 0.985 $\mu$m.  We have verified the wavelengths of both of these features using our three epochs of moderate resolution ($R\sim$1200) NIR SXD data.  In Figure \ref{fig:all_nir}, we show the SXD data obtained on 2010 Jan 3, covering the emission lines of \hei\ and [\ci] in the $J$-band.  

\section{Discussion} 

\subsection{The Nature of NIR Excess Emission in TWA 30B}

As noted, the shape of the 0.4--1.4 $\micron$ spectrum of TWA 30B is relatively steady and consistent with an M4 spectral classification while the 1.4--2.5 $\micron$ spectrum, in contrast, is highly variable.  This dichotomy suggests that the long wavelength flux originates from a distinct and variable component.  To isolate this NIR excess, we subtracted the M4 template Gl 213 from each epoch of data and fit a black body (BB) curve to the residuals.  BB fluxes were calculated for temperatures spanning 500-1500 K (in 10 K increments) and initially scaled to a source radius of 1 \rsun\ and a distance of 42 pc.  For each spectrum, we renormalized the BB fluxes to match the residual flux at 2.2 $\micron$.  To incorporate uncertainties in these spectral observations, we repeated the fits by adding a random Gaussian noise component pixel-by-pixel based on the noise spectrum.  This procedure was done 100 times with the set of BB curves fit to each iteration.  We then calculated the reduced $\chi^2$ values between the set of BB curves and the residuals over the wavelength ranges of 1.45--1.8 and 2.0--2.35 $\mu$m, excluding the region between these two bands affected by telluric absorption.  The best-fitting BB curve for this fit and the associated normalization were adopted as the temperature and visible area (in units \rsun$^2$), respectively, of the emitting region.  The adopted parameters (temperature and area) and uncertainties for each epoch are the averages and standard deviations for these simulated datasets.  These values are listed in Table \ref{table:specphot}; an example of a BB best-fit is shown in the right panel of Figure \ref{fig:all_nir}.

To interpret our results, we make a few assumptions about the origins of the infrared excess.  Given the good fit to a single temperature blackbody, we assume that it is reprocessed stellar radiation, emitted by some small (approximately single temperature) region of the disk\footnote{As such, we neglect the possibility that it is simply starlight scattered by the disk, since to zeroeth order this would not yield an excess spectral energy distribution mimicking a single temperature BB.}.  The BB temperature observed will then depend on both the radial distance from the star to the emitting region and the inclination of the region to the stellar photons.  Lacking {\it a priori} knowledge of the inclination, we first consider the two limiting cases: {\it (1)} an annulus in a thin flat disk (i.e., the emitting region is confined to the plane of the disk), and {\it (2)} a vertical disk rim or warp (i.e., the emitting region is perpendicular to the plane of the disk, therefore normal to the incident stellar radiation).  For a star of radius $R_{\ast}$ and effective temperature $T_{\ast}$, the temperature $T_d$ of a disk region at a radial distance $R_d$ for case {\it (1)} is given by:

$$T_d^4 = \frac{T_{\ast}^4}{\pi}\left[\frac{2}{3}\left(\frac{R_{\ast}}{R_d}\right)^3\right] \eqno[1]$$

\noindent which holds in the limit $R_d \gg R_{\ast}$ (see Equation [A22] in \citealt{1988ApJ...326..865A})\footnote{The exact equation -- [A22] in \cite{1988ApJ...326..865A}, retaining only the reprocessing term -- reads: ${T_d}^4 = ({T_{\ast}}^4/\pi) [{\rm arcsin}(u) - u(1-u^2)^{1/2}]$, where $u \equiv (R_{\ast}/R_d)$.  Using a Taylor expansion in the limit of $u \ll 1$ gives ${T_d}^4 = ({T_{\ast}}^4/\pi)(2u^3/3)$, which is our equation [1] above.  Comparing the value of $[{\rm arcsin}(u) - u(1-u^2)^{1/2}]$ to that of its Taylor expansion approximation $(2u^3/3)$, results in a difference of $<$5\% for $u \le 0.4$; i.e., the use of the approximation in equation [1] is valid, with very small errors, for any $R_d \ge 2.5 R_{\ast}$.}.

For case {\it (2)} the analogous expression is:

$$T_d^4 = {T_{\ast}^4}\left(\frac{R_{\ast}}{R_d}\right)^{2} \eqno[2]$$

\noindent Both equations contain the usual dilution factor of $(R_{\ast}/R_d)^2$.  The thin flat disk case differs in having an additional projection factor that goes to $R_{\ast}$/$R_d$ for large $R_d$ (see footnote [4]).  For inclinations intermediate between a flat disk and a vertical rim/warp, the disk temperature will vary approximately as $T_d \propto R_d^{-q}$, with $q$ ranging between the limiting values of 3/4 (eqn. [1]) and 1/2 (eqn. [2]).  Conversely, for a given BB temperature $T_d$, equations [1] and [2] provide the lower and upper limits, respectively, on the radius $R_d$ from which the emission can arise. 

Knowing the BB disk temperatures from our fits to the observed excess, we invert equations [1] and [2] to estimate the corresponding $R_d$ limits.  We assume $T_{\ast}$ = 3200K and $R_{\ast}$ = 1 R$_{\odot}$, appropriate for a young M4 star, using the spectral type to effective temperature conversion scale for young objects from \cite{2003ApJ...593.1093L} and evolutionary models from \cite{1998A&A...337..403B}.  These adopted values correspond to a mass of $\sim$0.15 M$_{\odot}$\ at an age of $\sim$10 Myr.  The inferred $R_d$ are listed in Table \ref{table:specphot}.  We see that, for a thin flat disk, $R_d$ is of order 3--5 $R_{\ast}$, while for a vertical rim or warp, the values are a factor of $\sim$5 higher\footnote{Note that the $R_d$ inferred for the thin flat disk case are all large enough -- $R_d > 2.5 R_{\ast}$ -- to be {\it a posteriori} consistent with the use of the approximate equation [1], as explained in footnote [4].}.  We note that the small $R_d$ values calculated for the thin flat disk case are consistent with the inner edge of a magnetospherically-truncated accretion disk, where disk structure is expected (see below).    

In both cases, a lower $T_d$ is associated with a larger $R_d$, which follows from our assumption that the radiation is reprocessed starlight from a disk region with a {\it fixed} inclination (either from a flat disk or a vertical rim/warp).  However, the $T_d$ variations may equally well arise from regions with {\it varying} inclination at roughly constant $R_d$, i.e., by changing $q$ instead of $R_d$ in the general reprocessing approximation $T_d \propto R_d^{-q}$.  For example, a rim or warp can form at the inner edge of the disk, due to frontal heating by stellar irradiation \citep{2001A&A...371..186N,2005ApJ...621..461D} and/or could be formed from the interaction between the stellar magnetosphere and the disk (\citealt{2000A&A...360.1031T,2003A&A...409..169B}; also invoked for TWA 30A: \citealt{2010ApJ...714...45L}).  So suppose we {\it assume} that the emission arises from a fixed radius at the disk inner edge, with the lowest observed $T_d$ arising from a flat disk geometry and higher values arising from a burgeoning warp, i.e., the emission region at larger inclinations.  Table \ref{table:specphot} shows that our smallest $T_d$ is consistent with flat disk emission from $R_d$ $\sim$ 5 $R_{\ast}$.  For $R_d$ fixed at this value, the highest $T_d$ would then require an inclination corresponding to $q$ $\sim$ 0.6, and the remaining $T_d$ would require $0.6<q<0.75$: in all cases intermediate between the flat and vertical values of 0.75 and 0.5, respectively.

So far we have discussed the geometrical constraints -- location and inclination -- on the emitting region implied by our inferred BB temperatures, assuming the excess flux is reprocessed stellar radiation from the disk.  In the next section (4.3), we present several lines of evidence suggesting that the disk around TWA30B is edge-on to our line of sight.  Therefore, the temporal variability in the NIR excess may be most straightforwardly interpreted as spatial variations in the disk structure rotating into our line of sight and out of it (occulted by the star and foreground parts of the disk).  If so, relatively smooth azimuthal variations in disk structure should cause variability on timescales comparable to the orbital period at the emitting radius.  For the stellar mass and radius assumed, the $R_d$ $\sim$ 3--5 $R_{\ast}$ derived for the thin flat disk case (and corresponding to the expected location of the disk inner edge) imply Keplerian orbital periods of $\sim$2--4 days (Table \ref{table:specphot}), consistent with the observed variations in the excess over timescales of a day.  The $R_d$ $\sim$ 15--25 $R_{\ast}$ derived for a purely vertical rim/warp, on the other hand, imply orbital periods of $\sim$15--40 days (Table \ref{table:specphot}), much longer than the observed timescales.  This suggests that the inner edge of the disk, where disk structure may naturally arise as noted above, may indeed be the source of the NIR excess.  We note that if an edge-on disk is significantly occulting the star (as argued in Section 4.3), then we are unlikely to be directly observing the inner regions of the disk where the NIR excess appears to arise (since the stellar occultation is more plausibly caused by the flared outer regions of the disk, which would occlude the inner disk as well).  In this case, it is likely that the inner disk emission is being scattered into our line of sight by surrounding disk regions (see also below).  
  
Lastly, we consider trends in the area of the emitting region inferred from our BB fits.  Fig. 6 shows the derived source areas plotted against the BB temperatures for all the observed SpeX prism spectra.  It is clear that the emitting area decreases with increasing temperature.  To explain this trend, we use the following salient points: the areas we derive do not correspond to the total area of the emitting region, but to the {\it fraction} of that area from which radiation actually reaches us; the variability timescale of the excess is consistent with emission from the disk inner edge; the derived $T_d$ are consistent with reprocessed emission from a region at fixed $R_d$ corresponding to this edge, but with variable inclination; and the observed emission is likely not reaching us directly, but via scattering off surrounding disk surfaces.  Now, for a fixed $R_d$, the associated $T_d$ increases with increasing inclination (i.e., as the source region becomes more vertical).  Thus the inverse trend in area with $T_d$ can be interpreted as a decrease in the apparent source area with increasing inclination, if indeed $R_d$ is roughly constant.  A more vertical region will emit a larger fraction of its radiation towards the central star, leaving less to be scattered by surrounding disk surfaces into our line of sight.  This would automatically yield a reduction in the apparent source area with increasing inclination (and hence with increasing $T_d$ for fixed $R_d$), as observed.    

The scenario postulated above -- reprocessed starlight emitted by disk structure close to the disk inner edge, and variably scattered towards us -- is clearly a simplistic toy model: while it is consistent with the temperatures, apparent source areas and variability timescales of the observed NIR excess, it is only loosely constrained so far by these parameters.  Further observations at higher temporal sampling, and extended to longer wavelengths are required to establish its veracity.    

\subsection{Comparison of TWA 30A and 30B}

Both components of this very wide ($\rho\sim$3400 AU) binary share many physical properties, such as a near-equivalent spectral type (M5~$\pm$~1 for TWA 30A and M4~$\pm$~0.5 for TWA 30B) and strong FELs of [\oi], [\oii], [\nii], and [\sii].  These FELs are spectroscopic signatures of an outflow.  In both cases the small velocity shifts ($\lesssim$30 \kms) of these lines from the rest velocity of the star suggest that the outflow lies perpendicular to the line-of-sight and, by extension, that the disk orientation lies close to edge-on.  The narrow P Cygni feature in the \nai D lines of both stars also points toward the outflow being viewed perpendicular to the line of sight.  The presence of [\oiii] in the spectrum of cTTS is rare and has only been previously reported in the literature in two high accretion rate cTTS -- DG Tau \citep{1995ApJ...452..736H,1996RMxAA..32..161R} and XZ Tau B \citep{2001ApJ...556..265W} -- and in TWA 30A (Looper et al. 2010).  This high-ionization line indicates collisional shock fronts (e.g., \citealt{1983ARA&A..21..209S,1995Ap&SS.233...11B}), typically seen in planetary nebulae or winds from massive stars, where the requisite extremely high temperatures and low densities can be found.  Given that the TWA 30AB system resides far from any molecular cloud \citep{2009PASJ...61..585T}, Looper et al. (2010) posit that this line may be formed by newer outflow material overtaking and collisionally shocking older jet fronts.  

With strong spectral signatures of outflows, it is at first surprising that the typical signatures of accretion in cTTS -- strong H$\alpha$ (usually defined as $>-$10 \AA) and \hei\ emission at 4471, 5876, 6678, and 7065 \AA\ -- are weak or absent in TWA 30A and TWA 30B.  The EWs of H$\alpha$ for both components is $-$7 \AA.  \cite{2010ApJ...714...45L} posit that H$\alpha$ emission, which in addition to chromospheric origins is formed in the shock region where infalling accreting gas impacts the stellar surface, is obscured by optically thick gas from a nearly edge-on disk.  \hei\ is also formed in the shock region, producing the narrow component, and in the accretion flow, producing the broad component.  TWA 30A was weakly detected in the RASS Faint Source Catalog as having X-ray emission while no such detection was made for TWA 30B.  An edge-on configuration could likewise obscure the X-ray emitting region of the star.  

Many of the emission lines, particularly FELs, are common to spectra of both stars, but a few differences exist.  \mgi] $\lambda$4571 emission appears in the spectrum of TWA 30A but not in TWA 30B.  The detection of this line is similar in rarity to [\oiii] emission in cTTS and is seen in only two other PMS stars -- XZ Tau B, which also has [\oiii] emission, and the EXor VY Tauri (M0; \citealt{1990ApJ...360..639H}).  This line indicates the presence of gas in the circumstellar disk.  The detection of \mgi] in higher mass cTTS remains prohibitively difficult due to the strong continuum from the star around this wavelength \citep{2008ApJ...688..398E}.  

[\ci] emission is present in both our optical and low-resolution NIR spectra in TWA 30B but is unseen in TWA 30A.   This line has been detected in the spectra of planetary nebulae \citep{1983ApJ...268..683J,1995MNRAS.273...47L}, molecular clouds \citep{1978A&A....68L...7H,1982NYASA.395..170M}, R Coronae Borealis stars \citep{2004MNRAS.353..143P}, and cometary tails \citep{2002ApJ...581..770O}.  The presence of [\ci] in the spectrum of a PMS star has not been previously reported in the literature.  The ratio of this line with the fine structure MIR lines of [\ci] offers a potentially powerful tool to constrain the size of the gaseous component of a protoplanetary disk \citep{2009A&A...496..725E}.  As of yet, no mid-infrared (MIR) observations of TWA 30B have been made.  

TWA 30A and TWA 30B present similar lines of evidence for a near edge-on orientation of an accreting circumstellar disk.  However, it is apparent that the disk of TWA 30B is viewed fully edge-on and is occulting the star.  This is highlighted from the fact that TWA 30B 
appears $\sim$5 mag underluminous in optical and NIR bands compared to TWA 30A despite being one spectral type earlier (Table \ref{table:TWA30ABproperties}).  TWA 30A has temporal changes in both emission line strengths profiles and variable reddening (A$_V$~=~1.5--9.0 mag), leading \cite{2010ApJ...714...45L} to suggest that the stellar axis may be inclined to the plane of the disk.  While we do not have the same coverage in the optical data of TWA 30B as obtained for TWA 30A, the two optical spectra of TWA 30B and the $J$-band portion of the 18 epochs of low-resolution NIR data show little variation in spectral slope or evidence for reddening.  Instead, the NIR excess present in the $H$- and $K$-band portions of the NIR spectral data show large variations between epochs on timescales as short as one day.  In Section 4.1, we proposed that these changes may arise from disk structure, such as a rim or a warp, at the inner disk edge that emits reprocessed starlight.

Assuming a distance of 42~$\pm$~2 pc, TWA 30A and TWA 30B would be the 2nd and 3rd closest accretors to the Sun.  The closest accreting star is another TWA system, the M4 PMS star TWA 3 (Hen 3-600; \citealt{1976ApJS...30..491H}), with a distance of 34~$\pm$~4 pc \citep{2005ApJ...634.1385M}.  Distances to the other nearby accreting TWA members are: TW Hya (53~$\pm$~3 pc), TWA 28 (55~$\pm$~1.5 pc), and TWA 27AB (53~$\pm$~1 pc), from Mamajek, in prep.  Both stars in the TWA 30AB system power the strongest jets of any TWA systems based on the strengths of their FELs.  Given the proximity of the system, they are excellent candidates for follow-up imaging to resolve their outflows.

\subsection{Are TWA 30AB Members of the $\beta$ Pictoris Group?}

While TWA 30A and TWA 30B have kinematics and spectroscopic indicators pointing towards membership in the TWA, there exists another notable young group of stars within tens of pc of the Sun -- the $\beta$ Pic moving group ($\sim$12 Myr; \citealt{2001ApJ...562L..87Z}). There has been some recent discussion on whether some previously claimed TWA stars could actually be $\beta$ Pic group members.  Most notably, the low-mass star TWA 22 \citep{2003ApJ...599..342S} was claimed to be the closest TWA member.  It was subsequently rejected as a TWA member by \cite{2005ApJ...634.1385M} based on its poor fit in the convergent point solution to the group.  This claim of non-membership was then refuted by \cite{2006ApJ...652..724S}.  Recently, \cite{2009A&A...503..281T} showed that TWA 22 is much more likely to be a $\beta$ Pic member than a TWA member based on a precise space velocity and kinematic traceback, enabled by a precise trigonometric parallax. The tale of the membership of ``TWA'' 22 reiterates that the space motions of nearby young groups are fairly similar at the \kms\ level, and membership may not be obvious.

Using the space motion for the $\beta$ Pic group from \cite{2004ARA&A..42..685Z}, we compared the proper motions and radial velocities for TWA 30A and TWA 30B to that expected for an ideal $\beta$ Pic member. The peculiar (perpendicular tangential) velocities for TWA 30A and TWA 30B would be 3.0~$\pm$~0.3 \kms\ and 1.6~$\pm$~1.1 \kms, respectively, and the predicted radial velocities would be 6.9 \kms\ each if $\beta$ Pic members.  In comparison, if TWA 30A and TWA 30B are members of the TWA, the peculiar velocities are only 1.2~$\pm$~0.3 \kms\ and 0.1~$\pm$~0.9 \kms, respectively.  The measured radial velocities of TWA 30A and TWA 30B are 12.3~$\pm$~1.5 \kms\ and 12~$\pm$~3 \kms, respectively; well in-line with the predicted radial velocity of 11.0~$\pm$~0.9 \kms\ for each if they are true TWA members.  Given the kinematic data in hand, we reject the hypothesis that the TWA 30 pair could be $\beta$ Pic members rather than TWA members.

\section{Conclusion}

We report the discovery and characterization of the M4 PMS star TWA 30B, which has a projected separation of $\sim$3400 AU from the M5 PMS star TWA 30A.  This binary is a member of the TWA and has stronger spectroscopic signatures of outflows than any other TWA members, with FELs of [\oi], [\oii], [\nii], and [\sii].  Both members display strong [\oiii] emission, which is rare in cTTS and possibly indicates collisional shock fronts from the jets.  All FELs have negligible radial velocity shifts from the stellar rest frame for both TWA 30A and 30B, suggesting that the outflows from both components are oriented in the plane of the sky.  TWA 30B is underluminous by $\sim$5 mag in both optical and NIR photometric bands in comparison to TWA 30A, suggesting that an edge-on disk heavily obscures the stellar photosphere.  TWA 30B has variable NIR excess well-fit by black body fluxes, with temperatures ranging from 630 to 880 K and emitting areas that are inversely proportional to the temperature.  The detection of [\ci] emission at 8727, 9824, and 9850 \AA\ mark the first detection of this line in a PMS star.  If paired with a future measurement of the [\ci] fine-structure lines in the MIR, these measurements could afford an independent measurement of the gaseous component of the circumstellar disk.  The TWA 30AB system represents an invaluable case study in low-mass stars as the 2nd and 3rd closest accretors to the Sun (42~$\pm$~2 pc), marked by a wide array of forbidden emission lines and temporal variability in NIR excess arising from each star's circumstellar disk.

\acknowledgments

DLL thanks Toby Owen and George Herbig for partial support during this research, John Rayner for advising, and Ben Zuckerman for helpful comments.  We thank our telescope operators at Magellan: Mauricio Martinez and Hern\'{a}n Nu\~{n}ez  and at IRTF: Dave Griep, Paul Sears, and Bill Golisch.  This research has benefitted from the M, L, and T dwarf compendium housed at DwarfArchives.org and maintained by Chris Gelino, Davy Kirkpatrick, and Adam Burgasser.  This research has made use of the Atomic Line List v2.04, maintained at \url{http://www.pa.uky.edu/$\sim$peter/atomic/}, the SIMBAD database and VizieR catalogue access tool, operated at CDS, Strasbourg, France, and the facilities of the Canadian Astronomy Data Centre operated by the National Research Council of Canada with the support of the Canadian Space Agency.  This publication makes use of data products from the Two Micron All Sky Survey, which is a joint project of the University of Massachusetts and the Infrared Processing and Analysis Center/California Institute of Technology, funded by the National Aeronautics and Space Administration and the National Science Foundation.  This research has made use of the NASA/ IPAC Infrared Science Archive, which is operated by the Jet Propulsion Laboratory, California Institute of Technology, under contract with the National Aeronautics and Space Administration.  As some spectroscopic follow-up data was obtained from the summit of Mauna Kea, the authors wish to recognize and acknowledge the very significant and cultural role and reverence that this mountaintop has always had with the indigenous Hawaiian community.  We are most fortunate to have the opportunity to conduct observations there.

Facilities: \facility{IRTF~(SpeX), Magellan:Clay~(MagE), Blanco~(ISPI)}

\clearpage

\begin{deluxetable}{llcclll}
\tablewidth{6.5in}
\tablenum{1}
\tablecaption{Astrometry for TWA 30B} 
\tablehead{ 
\colhead{RA (deg)} & \colhead {Dec (deg)} & \colhead{$\sigma_{RA}$ (mas)} &
\colhead{$\sigma_{Dec}$ (mas)} & \colhead{Epoch (yr)} & \colhead{Survey} & \colhead{Ref}}
\startdata
173.076795 & $-$30.308531 & 200 & 200 & 1968.7 & USNO-B1.0 & 1 \\ 
173.076125 & $-$30.308692 & 207 & 200 & 1994.1 & USNO-B1.0 & 1 \\
173.075928 & $-$30.308792 & 80 & 60 & 1999.23 & 2MASS & 2 \\
173.075736 & $-$30.308848 & 20 & 20 & 2008.95 & ISPI & 3 \\
173.075692 & $-$30.308861 & 20 & 20 & 2009.19 & ISPI & 3 \\
173.075672 & $-$30.308865 & 20 & 20 & 2009.92 & ISPI & 3 \\
173.075711 & $-$30.308872 & 20 & 20 & 2010.08 & ISPI & 3 \\
\enddata
\tablecomments{References: (1) USNO-B1.0 \citep{2003AJ....125..984M}, 
(2) 2MASS \citep{2003yCat.2246....0C, 2006AJ....131.1163S} and (3) this paper.}
\label{table:astrometry}
\end{deluxetable}

\begin{deluxetable}{lllllllll}
\tablewidth{7.0in}
\tablenum{2}
\tabletypesize{\scriptsize}
\tablecaption{Spectroscopic Observation Log for TWA 30A and 30B}
\tablehead{
\colhead{Object} & 
\colhead{Tel./Inst.} & 
\colhead{$\lambda$ ($\mu$m)} & 
\colhead{$\lambda$/$\Delta \lambda$} & 
\colhead{UT Date\tablenotemark{a}} & 
\colhead{N $\times$ t (s)\tablenotemark{b}} & 
\colhead{Z\tablenotemark{c}} & 
\colhead{Calibrator} & 
\colhead{Conditions}}
\startdata
TWA 30B & Magellan/MagE & 0.30--1.05 & 4100 & 090304 & 1~$\times$~1800 & 1.0 & Hiltner 600 (WD) & Clear, 0$\farcs$7 seeing \\
TWA 30B & Magellan/MagE & 0.30--1.05 & 4100 & 090530 & 1~$\times$~2400 & 1.1 & GD 108 (WD) & Thin Clouds, 0$\farcs$9 seeing \\
TWA 30B & IRTF/SpeX & 0.70--2.50 & 150 & 090514 & 8~$\times$~150 & 1.6 & HD 98949 (A0 V) & Clear, 0$\farcs$5 seeing \\
TWA 30B & IRTF/SpeX & 0.70--2.50 & 150 & 090628 & 4~$\times$~120 & 2.3 & HD 98949 (A0 V) & Clear, 0$\farcs$5 seeing \\
TWA 30B & IRTF/SpeX & 0.70--2.50 & 150 & 090629 & 4~$\times$~180 & 2.7 & HD 98949 (A0 V) & Clear, 0$\farcs$6 seeing \\
TWA 30B & IRTF/SpeX & 0.70--2.50 & 150 & 091228 & 10~$\times$~150 & 1.6 & HD 98949 (A0 V) & Clear, 0$\farcs$6 seeing \\
TWA 30B & IRTF/SpeX & 0.70--2.50 & 150 & 100102 & 10~$\times$~150 & 1.6 & HD 98949 (A0 V) & Clear, 0$\farcs$8 seeing \\
TWA 30B & IRTF/SpeX & 0.70--2.50 & 150 & 100103 & 10~$\times$~150 & 1.6 & HD 98949 (A0 V) & Clear, 0$\farcs$6 seeing \\
TWA 30B & IRTF/SpeX & 0.80--2.40 & 1200 & 100103 & 24~$\times$~180 & 1.6 & HD 98949 (A0 V) & Clear, 0$\farcs$6 seeing \\
TWA 30B & IRTF/SpeX & 0.70--2.50 & 150 & 100124 & 12~$\times$~150 & 1.6 & HD 98949 (A0 V) & Light Cirrus, 0$\farcs$6 seeing \\
TWA 30B & IRTF/SpeX & 0.80--2.40 & 1200 & 100124 & 26~$\times$~180 & 1.6 & HD 98949 (A0 V) & Light Cirrus, 0$\farcs$6 seeing \\
TWA 30B & IRTF/SpeX & 0.70--2.50 & 150 & 100125 & 16~$\times$~150 & 1.6 & HD 98949 (A0 V) & Clear, 0$\farcs$7 seeing \\
TWA 30B & IRTF/SpeX & 0.70--2.50 & 150 & 100126 & 20~$\times$~150 & 1.6 & HD 98949 (A0 V) & Clear, 0$\farcs$9 seeing \\
TWA 30B & IRTF/SpeX & 0.70--2.50 & 150 & 100127 & 10~$\times$~150 & 1.9 & HD 98949 (A0 V) & Clear, 0$\farcs$9 seeing \\
TWA 30B & IRTF/SpeX & 0.70--2.50 & 150 & 100128 & 18~$\times$~150 & 1.7 & HD 98949 (A0 V) & Clear, 0$\farcs$9 seeing \\
TWA 30B & IRTF/SpeX & 0.70--2.50 & 150 & 100520 & 10~$\times$~120 & 1.6 & HD 98949 (A0 V) & Clear, 1$\farcs$5 seeing \\
TWA 30B & IRTF/SpeX & 0.70--2.50 & 150 & 100521 & 12~$\times$~180 & 1.6 & HD 98949 (A0 V) & Clear, 0$\farcs$7 seeing \\
TWA 30B & IRTF/SpeX & 0.70--2.50 & 150 & 100522 & 6~$\times$~180 & 1.6 & HD 98949 (A0 V) & Clear, 0$\farcs$5 seeing \\
TWA 30B & IRTF/SpeX & 0.80--2.40 & 1200 & 100522 & 24~$\times$~180 & 1.7 & HD 98949 (A0 V) & Clear, 0$\farcs$5 seeing \\
TWA 30B & IRTF/SpeX & 0.70--2.50 & 150 & 100523 & 10~$\times$~180 & 1.8 & HD 98949 (A0 V) & Clear, 1$\farcs$5 seeing \\
TWA 30B & IRTF/SpeX & 0.70--2.50 & 150 & 100525 & 10~$\times$~180 & 1.6 & HD 98949 (A0 V) & Clear, 0$\farcs$6 seeing \\
TWA 30B & IRTF/SpeX & 0.70--2.50 & 150 & 100526 & 10~$\times$~180 & 1.6 & HD 98949 (A0 V) & Clear, 0$\farcs$7 seeing \\
TWA 30B & IRTF/SpeX & 0.70--2.50 & 150 & 100527 & 10~$\times$~180 & 1.6 & HD 98949 (A0 V) & Clear, 0$\farcs$7 seeing \\
\hline
TWA 30A & IRTF/SpeX & 0.80--2.40 & 1200 & 091229 & 10~$\times$~120 & 1.6 & HD 98949 (A0 V) & Clear, 0$\farcs$4 seeing \\
TWA 30A & IRTF/SpeX & 0.80--2.40 & 1200 & 091231 & 16~$\times$~120 & 1.7 & HD 98949 (A0 V) & Clear, 0$\farcs$3 seeing \\
TWA 30A & IRTF/SpeX & 0.80--2.40 & 1200 & 100102 & 6~$\times$~120 & 1.5 & HD 98949 (A0 V) & Clear, 0$\farcs$8 seeing \\
TWA 30A & IRTF/SpeX & 0.80--2.40 & 1200 & 100124 & 6~$\times$~120 & 1.8 & HD 98949 (A0 V) & Clear, 0$\farcs$6 seeing \\
TWA 30A & IRTF/SpeX & 0.80--2.40 & 1200 & 100125 & 22~$\times$~120 & 1.7 & HD 98949 (A0 V) & Clear, 0$\farcs$7 seeing \\
TWA 30A & IRTF/SpeX & 0.80--2.40 & 1200 & 100126 & 12~$\times$~120 & 1.8 & HD 98949 (A0 V) & Clear, 0$\farcs$8 seeing \\
TWA 30A & IRTF/SpeX & 0.80--2.40 & 1200 & 100520 & 10~$\times$~120 & 1.6 & HD 98949 (A0 V) & Clear, 1$\farcs$5 seeing \\
TWA 30A & IRTF/SpeX & 0.80--2.40 & 1200 & 100525 & 10~$\times$~120 & 1.6 & HD 98949 (A0 V) & Clear, 0$\farcs$9 seeing \\
\enddata
\tablenotetext{a}{Epochs are denoted as YYMMDD in all tables.}
\tablenotetext{b}{Number of integrations times the integration time.}
\tablenotetext{c}{Airmass.}
\label{table:log}
\end{deluxetable}

\begin{deluxetable}{lllll} 
\tablewidth{6.7in}
\tablenum{3}
\tablecaption{Properties of TWA 30AB}
\tablehead{ 
\colhead{Parameter} & \colhead{TWA 30A} & \colhead{TWA 30B} & \colhead{Adopted Mean} & \colhead{Ref}}
\startdata
2MASS Designation\tablenotemark{a} & J11321831$-$3019518 & J11321822$-$3018316 & \nodata & 1 \\
$\alpha$\degree (J2000)\tablenotemark{a} & 173.076322 & 173.075928 & \nodata & 1 \\
$\delta$\degree (J2000)\tablenotemark{a} & $-$30.331059 & $-$30.308792 & \nodata & 1 \\
\pmra~(\masyr) & $-$89.6~$\pm$~1.3 & $-$83~$\pm$~9 & \nodata & 2,3 \\
\pmdec~(\masyr) & $-$25.8~$\pm$~1.4 & $-$30~$\pm$~9 & \nodata & 2,3 \\
Distance\tablenotemark{b} (pc) & 42~$\pm$~2 & 44~$\pm$~5 & 42~$\pm$~2 & 3,4 \\
V$_{rad}$ (\kms) & 12.3~$\pm$~1.5 & 12~$\pm$~3 & 12~$\pm$~2  & 3,4 \\
Age (Myr) & $\sim$8 & $\sim$8 & $\sim$8 & 3,4 \\
Optical SpT & M5~$\pm$~1 & M4~$\pm$~0.5 & \nodata &  3,4 \\
$B$ (mag)\tablenotemark{c} & 15.6 & 20.4 & \nodata & 5 \\ 
$R$ (mag)\tablenotemark{c} & 12.9 & 17.4 & \nodata & 5 \\
$I$ (mag)\tablenotemark{c} & 11.30~$\pm$~0.03 & 16.68~$\pm$~0.08 & \nodata & 6 \\
DENIS $J$ (mag)\tablenotemark{c} & 9.78~$\pm$~0.06 & 14.00~$\pm$~0.09 & \nodata & 6 \\
DENIS $K$ (mag)\tablenotemark{c} & 8.88~$\pm$~0.07 & 11.08~$\pm$~0.08 & \nodata & 6 \\
DENIS $J-K$ (mag)\tablenotemark{c} & 0.90~$\pm$~0.09 & 2.92~$\pm$~0.12 & \nodata & 6\\
2MASS $J$ (mag)\tablenotemark{c} & 9.64~$\pm$~0.02 & 15.35~$\pm$~0.05 & \nodata & 1 \\
2MASS $H$ (mag)\tablenotemark{c} & 9.03~$\pm$~0.02 & 14.53~$\pm$~0.05 & \nodata & 1 \\
2MASS $K_s$ (mag)\tablenotemark{c} & 8.77~$\pm$~0.02 & 13.72~$\pm$~0.04 & \nodata & 1 \\
2MASS $J-K_s$ (mag)\tablenotemark{c} & 0.87~$\pm$~0.03 & 1.63~$\pm$~0.06 & \nodata & 1 \\
Angular Separation\tablenotemark{b} ($\arcsec$) & \nodata & \nodata & 80.2 & 3 \\
Projected Separation\tablenotemark{b} (AU) & \nodata & \nodata & 3370~$\pm$~160 & 3 \\
Est.\ Actual Separation\tablenotemark{b} (AU) & \nodata & \nodata & 4250~$\pm$~200 & 3 \\
Li EW (\AA) & 0.6~$\pm$~0.1 & 0.5~$\pm$~0.1\tablenotemark{d} & \nodata & 3,4 \\
H$\alpha$ EW (\AA) & $-$6.8~$\pm$~1.2 & $-$7.4~$\pm$~0.02\tablenotemark{e} & \nodata & 3,4 \\
\enddata
\tablenotetext{a}{From the 2MASS Point Source Catalog at epoch 24 March 1999 UT.}
\tablenotetext{b}{Estimated based on a convergent point analysis (see Section 3.1).}
\tablenotetext{c}{TWA 30A and TWA 30B have time variable photometry (Looper et al. 2010 and this paper).}  
\tablenotetext{d}{Derived from the equivalent width measured from the MagE data of epoch 2009 May 30 reported in Table \ref{table:EWs}.  We did not detect Li I in the MagE data of epoch 2009 March 4.}
\tablenotetext{e}{Derived from the average equivalent widths measured from the two MagE spectra reported in Table \ref{table:EWs}.}
\tablecomments{References: (1) 2MASS \citep{2006AJ....131.1163S}, (2) UCAC3 \citep{2010AJ....139.2200F}, 
(3) this paper, (4) \cite{2010ApJ...714...45L}, (5) USNO-A2.0 \citep{1998yCat.1252....0M}, and (6) DENIS \citep{1997Msngr..87...27E}.
}
\label{table:TWA30ABproperties}
\end{deluxetable}

\begin{deluxetable}{llllllc}
\tablewidth{4.8in}
\tablenum{4}
\tabletypesize{\scriptsize}
\tablecaption{Equivalent Widths for Selected Line Features of TWA 30B}
\tablehead{ 
\colhead{UT} & \colhead{$\lambda_{lab}$ (\AA)} & \colhead{$\lambda_{obs}$ (\AA)} & \colhead{Ion} & \colhead{Flux\tablenotemark{a}} & \colhead{EW~$\pm$~1$\sigma$ (\AA)\tablenotemark{b}} & \colhead{V (km s$^{-1}$)\tablenotemark{c}}}
\startdata
090304 & 3934.777 & 3935.06 & Ca II K & \nodata & $-$24.5 & 22 \\
090530 & 3934.777 & 3934.93 & Ca II K & \nodata & $-$104 &  12 \\
\hline
090304 & 3969.591 & 3969.90 & Ca II H + H$\epsilon$\tablenotemark{d} & \nodata & $-$3.3 & 23 \\
090530 & 3969.591 & 3969.73 & Ca II H + H$\epsilon$\tablenotemark{d} & \nodata & $-$25.1 & 11 \\
\hline
090304 & 4069.749 & 4070.029 & [S II] & 66.2 & $-$61~$\pm$~10 & 21 \\
090530 & 4069.749 & 4069.95 & [S II] & \nodata & $-$19.2 & 15 \\
\hline
090304 & 4077.500 & 4077.70 & [S II] & \nodata & $-$14.5 & 15 \\
090530 & 4077.500 & 4077.73 & [S II] & \nodata & $-$1.9 & 17 \\
\hline
090304 & 4862.683 & 4862.935 & H$\beta$ & 11.4 & $-$8~$\pm$~2 & 16 \\
090530 & 4862.683 & 4862.90 & H$\beta$ & \nodata & $-$7~$\pm$~2 & 13 \\
\hline
090304 & 4960.300 & 4960.765 & [O III] & 24.9 & $-$19~$\pm$~1 & 28 \\
090530 & 4960.300 & 4960.703 & [O III] & 24.8 & $-$19~$\pm$~3 & 24 \\
\hline
090304 & 5008.240 & 5008.465 & [O III] & 103 & $-$94~$\pm$~2 & 13 \\
090530 & 5008.240 & 5008.755 & [O III] & 81.2 & $-$72~$\pm$~3 & 31 \\
\hline
090304 & 5200.621 & 5200.438 & [Fe II] & 11.9 & $-$7~$\pm$~2 & $-$11 \\
090530 & 5200.621 & \nodata & [Fe II] & \nodata & $<-4$ & \nodata \\
\hline
090304 & 5578.887 & 5579.013 & [O I] & 11.0 & $-$6.5~$\pm$~0.4 & 7 \\
090530 & 5578.887 & 5579.000 & [O I] & 4.6 & $-$1.4~$\pm$~0.6 & 6 \\
\hline
090304 & 5756.240 & 5756.678 & [N II] & 10.6 & $-$4.2~$\pm$~0.7 & 23 \\
090530 & 5756.240 & \nodata & [N II] & \nodata & $<-$2 & \nodata \\
\hline
090304 & 5891.583 & 5892.409 & Na I D & 21.3 & $-$18.1~$\pm$~0.8 & 42 \\
090530 & 5891.583 & 5891.65 & Na I D & \nodata & $-$3.8 & 3 \\
\hline
090304 & 5897.558 & 5898.340 & Na I D & 18.2 & $-$15.1~$\pm$~0.8 & 40 \\
090530 & 5897.558 & 5897.67 & Na I D & \nodata & $-$2.6 & 6 \\
\hline
090304 & 6302.050 & 6302.283 & [O I] & 233 & $-$226.3~$\pm$~0.4 & 11 \\
090530 & 6302.050 & 6302.198 & [O I] & 92.0 & $-$84.7~$\pm$~0.6 & 7 \\
\hline
090304 & 6365.540 & 6365.789 & [O I] & 55.8 & $-$50.1~$\pm$~0.3 & 12 \\
090530 & 6365.540 & 6365.696 & [O I] & 26.5 & $-$20.9~$\pm$~0.1 & 7 \\
\hline
090304 & 6549.850 & 6550.107 & [N II] & 13.0 & $-$7.2~$\pm$~0.2 & 12 \\
090530 & 6549.850 & 6550.071 & [N II] & 8.9 & $-$3.0~$\pm$~0.1 & 10 \\
\hline
090304 & 6564.610 & 6564.779 & H$\alpha$ & 14.4 & $-$8.1~$\pm$~0.2 & 8 \\
090530 & 6564.610 & 6564.738 & H$\alpha$ & 13.3 & $-$7.2~$\pm$~0.6 & 6 \\
\hline
090304 & 6585.280 & 6585.526 & [N II] & 31.5 & $-$25.1~$\pm$~0.2 & 11 \\
090530 & 6585.280 & 6585.417 & [N II] & 17.6 & $-$10.8~$\pm$~0.1 & 6 \\
\hline
090304 & 6709.660 & \nodata & Li I & \nodata & $<$0.2 & \nodata \\
090530 & 6709.660 & 6709.831 & Li I & 3.0 & 0.46~$\pm$~0.13 & 8 \\
\hline
090304 & 6718.290 & 6718.528 & [S II] & 10.2 & $-$5.8~$\pm$~0.2 & 11 \\
090530 & 6718.290 & 6718.404 & [S II] & 7.4 & $-$2.9~$\pm$~0.2 & 5 \\
\hline
090304 & 6732.670 & 6732.909 & [S II] & 16.4 & $-$10.4~$\pm$~0.2 & 11 \\
090530 & 6732.670 & 6732.891 & [S II] & 11.4 & $-$6.0~$\pm$~0.2 & 10 \\
\hline
090304 & 7322.010 & 7322.258 & [O II] & 18.8 & $-$10.7~$\pm$~0.1 & 10 \\
090530 & 7322.010 & 7322.316 & [O II] & 12.7 & $-$4.7~$\pm$~0.1 & 13 \\
\hline
090304 & 7331.680 & 7332.663 & [O II] & 15.4 & $-$7.8~$\pm$~0.1 & 40 \\
090530 & 7331.680 & 7332.495 & [O II] & 10.5 & $-$3.4~$\pm$~0.1 & 33 \\
\hline
090304 & 7667.021 & 7667.435 & K I & 14.5 & $-$6.1~$\pm$~0.4 & 16 \\
090530 & 7667.021 & \nodata & K I & \nodata & \nodata & \nodata \\
\hline
090304 & 7701.093 & 7701.61 & K I & \nodata & $-$3.1 & 20 \\
090530 & 7701.093 & 7701.335 & K I & 3.5 & 0.46~$\pm$~0.07 & 9 \\
\hline
090304 & 8185.505 & 8185.537 & Na I & 4.1 & 0.76~$\pm$~0.12 & 1 \\
090530 & 8185.505 & 8185.800 & Na I & 3.6 & 0.66~$\pm$~0.09 & 11 \\
\hline
090304 & 8500.36 & \nodata & Ca II & \nodata & $<$0.2 & \nodata \\
090530 & 8500.36 & \nodata & Ca II & \nodata & $<$0.2 & \nodata \\
\hline
090304 & 8544.44 & 8545.08 & Ca II & \nodata & 0.52 & 22 \\
090530 & 8544.44 & 8545.32 & Ca II & \nodata & 0.86 & 31 \\
\hline
090304 & 8664.52 & 8664.73 & Ca II & \nodata & 0.62 & 7 \\
090530 & 8664.52 & 8665.07 & Ca II & \nodata & 0.73 & 19 \\
\hline
090304 & 8729.520 & 8729.531 & [C I] & 11.4 & $-$2.4~$\pm$~0.1 & 0 \\
090530 & 8729.520 & 8729.865 & [C I] & 6.8 & $-$0.4~$\pm$~0.1 & 12 \\
\hline
090304 & 9826.820 & 9826.803 & [C I] & 5.6 & $-$1.3~$\pm$~0.1 & $-$1 \\
090530 & 9826.820 & 9827.020 & [C I] & 4.9 & $-$0.6~$\pm$~0.1 & 6 \\
\hline
090304 & 9852.960 & 9853.173 & [C I] & 10.7 & $-$3.4~$\pm$~0.1 & 6 \\
090530 & 9852.960 & 9852.925 & [C I] & 6.4 & $-$1.4~$\pm$~0.1 & $-$1 \\
\enddata
\tablenotetext{a}{The integrated line fluxes are given in units of 10$^{-16}$ erg cm$^{-2}$ s$^{-1}$ and should only be used to calculate relative line fluxes between features in the same spectrum as our data are not photometrically calibrated and hence do not account for slit losses or non-photometric conditions.}
\tablenotetext{b}{Some lines were measured by hand and have typical uncertainties of 0.5--3 \AA.}
\tablenotetext{c}{Reported as heliocentric velocities, i.e., not relative to the assumed stellar RV of 12~$\pm$~3 \kms.}
\tablenotetext{d}{This line is blended or contaminated.}
\label{table:EWs}
\end{deluxetable}

\begin{deluxetable}{lllllllllllll}
\tablewidth{7.2in}
\tablenum{5}
\tabletypesize{\scriptsize}
\tablecaption{Derived NIR Magnitudes\tablenotemark{a}, Colors\tablenotemark{a}, and Disk Parameters of TWA 30AB}
\tablehead{ 
\colhead{Object} & 
\colhead{UT\tablenotemark{b}} & 
\colhead{$J$\tablenotemark{c}} & 
\colhead{$J-H$\tablenotemark{c}} & 
\colhead{$H-K_s$\tablenotemark{c}} & 
\colhead{$J-K_s$\tablenotemark{c}} & 
\colhead{BB ($T_d$)\tablenotemark{d}} & 
\colhead{Area\tablenotemark{e}} & 
\colhead{R$_{d1}$\tablenotemark{f,g}} & 
\colhead{P$_1$\tablenotemark{f,h}} & 
\colhead{R$_{d2}$\tablenotemark{f,g}} & 
\colhead{P$_2$\tablenotemark{f,h}} & 
\colhead{Notes}}
\startdata
TWA 30B & 990324 & 15.35 & 0.82 & 0.81 & 1.63 & \nodata & \nodata & \nodata & \nodata & \nodata & \nodata & 2MASS PSC \\
TWA 30B & 090514 & 15.4 & 0.70 & 0.69 & 1.39 & 670~$\pm$~20 & 4.4~$\pm$~0.7 & 4.8 & 3.2 & 22.8 & 33.0 & SpeX prism \\
TWA 30B & 090628 & 15.3 & 0.72 & 0.50 & 1.22 & 880~$\pm$~40 & 0.45~$\pm$~0.09 & 3.3 & 1.8 & 13.2 & 14.6 & SpeX prism \\
TWA 30B & 090629 & 15.3 & 0.73 & 0.56 & 1.29 & 860~$\pm$~40 & 0.48~$\pm$~0.10 & 3.4 & 1.9 & 13.8 & 15.6 & SpeX prism \\
TWA 30B & 091228 & 15.3 & 0.69 & 0.59 & 1.28 & 720~$\pm$~20 & 3.1~$\pm$~0.4 & 4.4 & 2.8 & 19.8 & 26.6 & SpeX prism \\
TWA 30B & 100102 & 15.1 & 0.82 & 1.06 & 1.88 & 703~$\pm$~8 & 3.8~$\pm$~0.2 & 4.5 & 2.9 & 20.7 & 28.5 & SpeX prism \\
TWA 30B & 100103 & 15.3 & 0.95 & 1.29 & 2.24 & 743~$\pm$~6 & 2.77~$\pm$~0.12  & 4.2 & 2.6 & 18.5 & 24.2 & SpeX prism \\
TWA 30B & 100124 & 15.2 & 0.66 & 0.50 & 1.16 & 630~$\pm$~50 & 8~$\pm$~4 & 5.2 & 3.6 & 25.8 & 39.7 & SpeX prism \\
TWA 30B & 100125 & 15.4 & 0.68 & 0.50 & 1.18 & 750~$\pm$~20 & 2.6~$\pm$~0.4 & 4.1 & 2.5 & 18.2 & 23.5 & SpeX prism \\
TWA 30B & 100126 & 15.2 & 0.65 & 0.47 & 1.12 & 710~$\pm$~30 & 3.7~$\pm$~0.7 & 4.4 & 2.8 & 20.3 & 27.7 & SpeX prism \\
TWA 30B & 100127 & 15.0 & 0.67 & 0.48 & 1.15 & 760~$\pm$~50 & 3.0~$\pm$~0.9 & 4.1 & 2.5 & 17.7 & 22.6 & SpeX prism \\
TWA 30B & 100128 & 15.3 & 0.67 & 0.46 & 1.13 & 780~$\pm$~30 & 2.3~$\pm$~0.4 & 3.9 & 2.3 & 16.8 & 20.9 & SpeX prism \\
TWA 30B & 100520 & 15.4 & 0.78 & 0.84 & 1.62 & 700~$\pm$~30 & 3.5~$\pm$~0.8 & 4.5 & 2.9 & 20.9 & 28.9 & SpeX prism \\
TWA 30B & 100521 & 15.6 & 0.80 & 0.82 & 1.62 & 790~$\pm$~30 & 1.9~$\pm$~0.3 & 3.9 & 2.3 & 16.4 & 20.1 & SpeX prism \\
TWA 30B & 100522 & 15.5 & 0.78 & 0.82 & 1.60 & 770~$\pm$~10 & 2.1~$\pm$~0.2 & 4.0 & 2.4 & 17.3 & 21.7 & SpeX prism \\
TWA 30B & 100523 & 15.1 & 0.77 & 0.81 & 1.58 & 760~$\pm$~20 & 2.8~$\pm$~0.4 & 4.1 & 2.5 & 17.7 & 22.6 & SpeX prism \\
TWA 30B & 100525 & 16.0 & 0.88 & 1.02 & 1.90 & 800~$\pm$~20 & 1.5~$\pm$~0.1  & 3.8 & 2.2 & 16.0 & 19.4 & SpeX prism \\
TWA 30B & 100526 & 15.8 & 0.86 & 1.12 & 1.98 & 758~$\pm$~12 & 2.0~$\pm$~0.2  & 4.1 & 2.5 & 17.8 & 22.8 & SpeX prism \\
TWA 30B & 100527 & 15.8 & 0.87 & 1.08 & 1.95 & 740~$\pm$~10 & 2.2~$\pm$~0.2 & 4.2 & 2.6 & 18.7 & 24.5 & SpeX prism \\
\hline
TWA 30A & 091227 & 9.9   & 0.72 & 0.33 & 1.05 & \nodata & \nodata & \nodata & \nodata & \nodata & \nodata & SpeX sxd \\
TWA 30A & 091231 & 10.1 & 0.75 & 0.35 & 1.10 & \nodata & \nodata & \nodata & \nodata & \nodata & \nodata & SpeX sxd \\
TWA 30A & 100102 & 9.9   & 0.70 & 0.32 & 1.02 & \nodata & \nodata & \nodata & \nodata & \nodata & \nodata & SpeX sxd \\
TWA 30A & 100124 & 10.0 & 0.72 & 0.33 & 1.05 & \nodata & \nodata & \nodata & \nodata & \nodata & \nodata & SpeX prism \\
TWA 30A & 100125 & 9.6   & 0.72 & 0.37 & 1.09 & \nodata & \nodata & \nodata & \nodata & \nodata & \nodata & SpeX sxd \\
TWA 30A & 100126 & 10.0 & 0.66 & 0.29 & 0.95 & \nodata & \nodata & \nodata & \nodata & \nodata & \nodata & SpeX sxd \\
TWA 30A & 100520 & 10.8 & 1.04 & 0.50 & 1.54 & \nodata & \nodata & \nodata & \nodata & \nodata & \nodata & SpeX sxd \\
TWA 30A & 100525 & 12.2 & 1.53 & 0.84 & 2.37 & \nodata & \nodata & \nodata & \nodata & \nodata & \nodata & SpeX sxd \\
\enddata
\tablenotetext{a}{All magnitudes and colors are derived using 2MASS filters for spectrophotometry on the NIR spectra taken in prism mode (as indicated) with the exception of the epoch marked '2MASS PSC,' which is photometry.  In photometric conditions, the derived {\it absolute} spectrophotometry, flux calibrated with nearby A0 V stars, are accurate to $\sim$10\%, while the derived {\it relative} spectrophotometry (colors) are accurate to a few percent \citep{2009ApJS..185..289R}.   For a full description, see Section 3.3.1.}
\tablenotetext{b}{UT dates are recorded as YYMMDD.}
\tablenotetext{c}{Magnitudes.}
\tablenotetext{d}{Temperatures (in K) of the NIR excess measured by fitting BB curves to the residuals of the M4 template Gl 213 subtracted from the TWA 30B data.}
\tablenotetext{e}{The apparent area of the emitting region, in units of \rsun$^2$.}
\tablenotetext{f}{Values denoted by "1" are derived from Equation [1] (the annulus of a thin disk case) while those denoted by "2" are derived from Equation [2] (the vertical rim/disk warp case).}
\tablenotetext{g}{The radial distance to the emitting black body flux arising from the disk, in units of \rsun.}
\tablenotetext{h}{The Keplerian rotation periods for each case, in units of days.}
\label{table:specphot}
\end{deluxetable}

\clearpage

\begin{figure*}[htbp]
\centering
\includegraphics[scale=0.8]{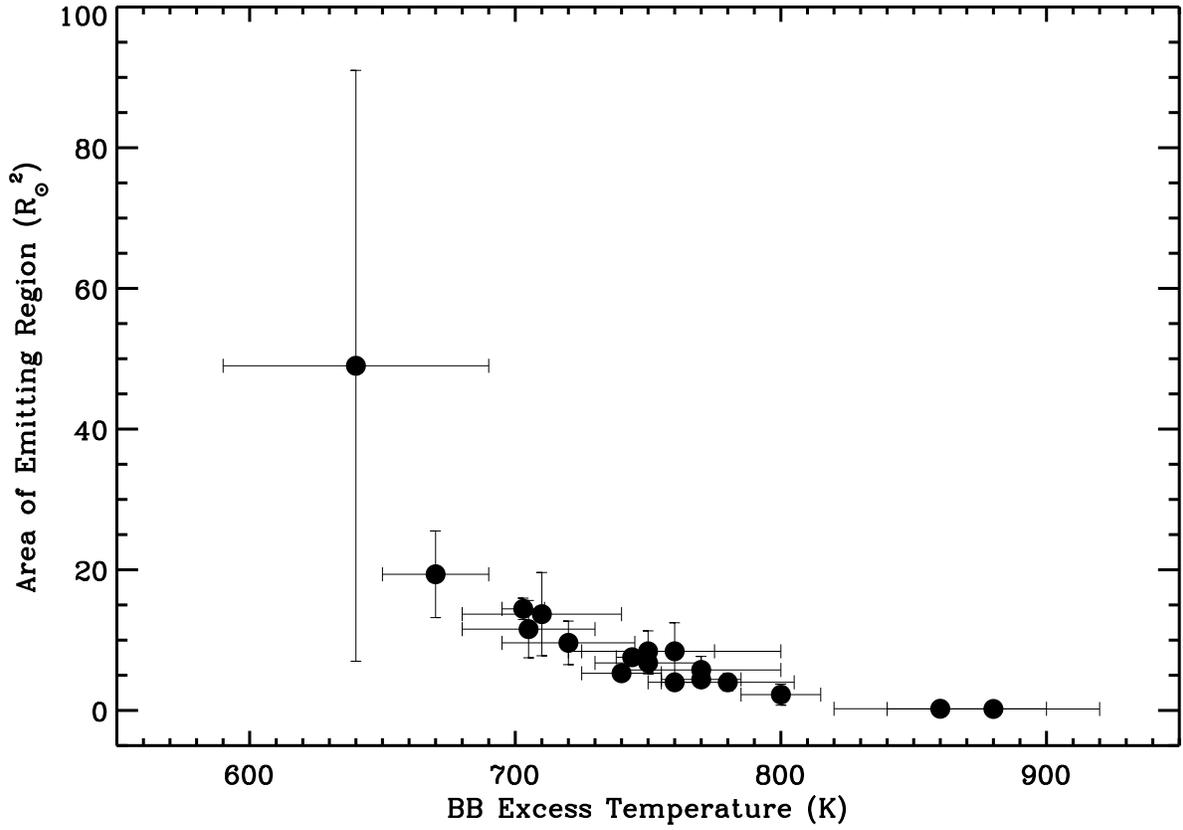}
\caption{The calculated area of the emitting flux region in units of \rsun$^2$ as a function of the blackbody (BB) temperatures calculated from the excess emission for each set of data.  To calculate the size of the emitting region we assumed that for the scaling factor (R/d)$^2$, R~=~1 \rsun\ and d~=~42 pc.  For further details, see Section 4.1.}
\label{fig:Rdist}
\end{figure*}

\clearpage


\end{document}